\documentclass[fleqn,usenatbib]{mnras}

\usepackage{newtxtext,newtxmath}

\usepackage[T1]{fontenc}

\DeclareRobustCommand{\VAN}[3]{#2}
\let\VANthebibliography\thebibliography
\def\thebibliography{\DeclareRobustCommand{\VAN}[3]{##3}\VANthebibliography}

\usepackage{graphicx}	
\usepackage{amsmath}	
\usepackage{diagbox}

\newcommand{\car}{^{12}\text{C}}
\newcommand{\oxy}{^{16}\text{O}}
\newcommand{\neo}{^{20}\text{Ne}}
\newcommand{\magn}{^{24}\text{Mg}}
\newcommand{\sil}{^{28}\text{Si}}

\title[Shell mergers in massive stars]{Shell mergers in 
the late stages of massive star evolution: 
new insight from 3D hydrodynamic simulations}

\author[Rizzuti et al.]{F. Rizzuti$^{1,2}$\thanks{E-mail: f.rizzuti@keele.ac.uk, federico.rizzuti@inaf.it}, R. Hirschi$^{1,3}$, V. Varma$^{1}$, W. D. Arnett$^{4}$, C. Georgy$^{5}$, C. Meakin$^{6}$, M. Mocák$^{1}$, A. StJ. Murphy$^{7}$ \newauthor and T. Rauscher$^{8,9}$
\\
$^{1}$Astrophysics Group, Lennard-Jones Laboratories, Keele University, Keele ST5 5BG, UK\\
$^{2}$Dipartimento di Fisica, Università degli Studi di Trieste, Via Tiepolo 11, I-34143 Trieste, Italy\\
$^{3}$Kavli IPMU (WPI), University of Tokyo, 5-1-5 Kashiwanoha, Kashiwa 277-8583, Japan\\
$^{4}$Steward Observatory, University of Arizona, 933 N. Cherry Avenue, Tucson AZ 85721, USA\\
$^{5}$Geneva Observatory, Geneva University, CH-1290 Sauverny, Switzerland\\
$^{6}$Pasadena Consulting Group, 1075 N Mar Vista Ave, Pasadena, CA 91104 USA\\
$^{7}$School of Physics and Astronomy, University of Edinburgh, Edinburgh EH9 3FD, UK\\
$^{8}$Department of Physics, University of Basel, Switzerland\\
$^{9}$Centre for Astrophysics Research, University of Hertfordshire, UK
}

\date{Accepted 2024 July 19. Received 2024 July 19; in original form 2024 May 4}

\pubyear{2024}

\begin{document}
\label{firstpage}
\pagerange{\pageref{firstpage}--\pageref{lastpage}}
\maketitle

\begin{abstract}
One-dimensional (1D) stellar evolution models are widely used across various astrophysical fields, however they are still dominated by important uncertainties that deeply affect their predictive power. Among those, the merging of independent convective regions is a poorly understood phenomenon predicted by some 1D models but whose occurrence and impact in real stars remain very uncertain. Being an intrinsically multi-D phenomenon, it is challenging to predict the exact behaviour of shell mergers with 1D models. In this work, we conduct a detailed investigation of a multiple shell merging event in a 20 M$_\odot$ star using 3D hydrodynamic simulations. Making use of the active tracers for composition and the nuclear network included in the 3D model, we study the merging not only from a dynamical standpoint but also considering its nucleosynthesis and energy generation. Our simulations confirm the occurrence of the merging also in 3D, but reveal significant differences from the 1D case. Specifically, we identify entrainment and the erosion of stable regions as the main mechanisms that drive the merging, we predict much faster convective velocities compared to the mixing-length-theory velocities, and observe multiple burning phases within the same merged shell, with important effects for the chemical composition of the star, which presents a strongly asymmetric (dipolar) distribution. We expect that these differences will have important effects on the final structure of massive stars and thus their final collapse dynamics and possible supernova explosion, subsequently affecting the resulting nucleosynthesis and remnant.
\end{abstract}

\begin{keywords}
convection - hydrodynamics - nuclear reactions, nucleosynthesis, abundances - stars: evolution –stars: interiors – stars: massive
\end{keywords}



\section{Introduction}

Stellar evolution is normally represented as a sequence of burning phases distributed over time. While this is accurate for describing the core burning phases, the question is more complex when studying the burning shells of massive stars. In contrast to the traditional view of an `onion-ring' structure, i.e.\ a system of concentric burning shells surrounding the core (see e.g.\ \citealp{1982phyn.book.....S}), 1D stellar evolution models have been showing that the occurrence and location of the burning shells depend on the properties of each star (mass, metallicity, rotation, overshoot), predicting that burning shells may appear, disappear, and reappear in the same or in a different location, with the same or with a different type of burning (see e.g.\ \citealp{Hirschi04,Sukhbold_2014}). Without computing the stellar model, one cannot predict exactly how a specific star would behave. \\
To further complicate the issue, 1D stellar evolution simulations may show another type of occurrence in the evolution of a star: the merging of multiple convective shells into a single convective one \citep{Rauscher_2002,Tur_2007}. This environment is very challenging to study, due to the complex interaction between convection, nuclear burning and entrainment, resulting in new dynamics and alternative nucleosynthesis paths that are difficult to include in 1D models with simplifying prescriptions. These effects are expected to have an important impact on the structure and chemical composition of the star, changing its stratification by the time of the collapse and its abundances due to convective mixing, therefore affecting also the possible supernova explosion and the chemical enrichment of the interstellar medium. \\
A major source of uncertainty concerning the formation and evolution of shell mergers is the limited literature investigating these episodes. 1D stellar evolution models have been reporting the occurrence of convective shell mergers for a long time: for example, \cite{Rauscher_2002} and \cite*{Tur_2007} observed the merging of convective oxygen-, neon-, and carbon-burning shells about 1 day before collapse in 20 M$_\odot$ models (but not in 19 and 21 M$_\odot$ models). More generally, \cite{Sukhbold_2014} found merged O, Ne and C-burning shells during the final hours of $15\text{ - }20 \text{ M}_\odot$ stars in their grid of non-rotating, solar metallicity models. \cite{10.1093/mnras/stx2470} have analysed an even larger grid of non-rotating, solar-metallicity models, and found a large prevalence of O, Ne and C-shell merging events above 15 M$_\odot$, with oxygen- and neon-burning in the same convective zone for 40 per cent of the stars between 16 - 26 M$_\odot$. More recently, \cite{2024ApJS..270...28R} found that stellar rotation favours C-O shell mergers in low-metallicity 15 M$_\odot$ models. These studies show how common these events are expected to be during the late phases of massive star evolution according to 1D models.\\
These works also suggest that shell-merging events in advanced phases of massive stars can be responsible for the production of isotopes that are difficult to explain otherwise, both during the convective phases and later on in the supernova explosions. Indeed, the merging event can transport elements to deeper regions where they can burn more rapidly, and at the same time bring the ashes closer to the surface, where they are more likely to be ejected into the interstellar medium. \\ 
In particular, carbon-oxygen merging shells are shown to be a potential main source for the nucleosynthesis of the odd-Z elements $^{31}$P, $^{35}$Cl, $^{39}$K, and $^{45}$Sc \citep{Rauscher_2002,10.1093/mnrasl/slx126}, whose production is currently underestimated by Galactic chemical evolution models (see e.g.\ \citealp{Cescutti12} for the origins of phosphorus). This happens because the heating of ingested carbon at oxygen-burning temperatures can trigger a sequence of $\gamma$-reactions that release free protons and produce the odd-Z elements; this is sometimes called ``p-process''. Some studies however suggest that the p-nuclei produced in this way are completely reprocessed during the passage of the following supernova shock \citep[see][]{1978ApJS...36..285W}. Additionally, during the explosive nucleosynthesis in core-collapse supernovae, the collapse of the star can trigger the photodisintegration of heavy isotopes across the carbon-oxygen merger site, causing the production of rare proton-rich isotopes beyond iron through a chain of photodisintegrations, also known as ``$\gamma$-process'' \citep[see][]{Rauscher_2002,roberti2023gammaprocess}.\\ 
Recently, some works also started to study these peculiar events with hydrodynamic models. \cite{10.1093/mnrasl/slx126} were among the first who performed 3D hydrodynamic simulations of carbon ingestion from a stable layer into a convective oxygen-burning shell, based on a stratification assumed from a 25 M$_\odot$ stellar evolution model. The resulting nucleosynthesis, that they computed with a 1D model based on the 3D simulations, confirmed that the high entrainment rates boost the production of the odd-Z elements $^{31}$P, $^{35}$Cl, $^{39}$K, and $^{45}$Sc through $(\gamma, \text{p})$ reactions. Their study of the consequent explosive nucleosynthesis also shows that the overproduction factors for these elements are little affected, indicating that their principal production sites are likely to be the convective merging shells. \\ 
Following the same approach, \cite{Andrassy2020} further investigated the ingestion of carbon into a convective oxygen-burning shell using 3D simulations that include explicit carbon- and oxygen-burning reactions. As a result, in addition to measuring an entrainment rate that can explain the production of the odd-Z elements, they estimated that the carbon-burning inside the oxygen shell can contribute to around 14 - 33 per cent of the total luminosity of the shell, showing how impactful the extra burning can be.\\
Finally, \cite{Mocak2018} have studied the ingestion of neon into a convective oxygen-burning shell for a 23 M$_\odot$ star, with 3D simulations in spherical geometry. For a more realistic scenario, they included an explicit 25-isotope network to reproduce the energy release dominated by oxygen- and neon-burning inside the convective shell. More specifically, neon burning results from the heating of entrained material into the convective layers, while oxygen burning is enhanced by the additional fuel from the stable regions. As a result, a new quasi-steady state is reached: two burning shells are present within the same convective zone, characterized by two distinct peaks in nuclear energy generation.\\
While these studies have been innovative, they have all focused their attention on environments where an alternative fuel is ingested from a stable region into a convective one. However, stellar simulations run with 1D models also show a rather different type of occurrence, i.e.\ the merging of multiple convective shells of different compositions. In such models, usually the neon or oxygen convective regions grow over time and eventually make contact with the shells above. The evolution of convective shells that begin their life separately and later merge is an extremely interesting environment to study with multi-D simulations, both for the effects of these extreme dynamics on the stellar structure, and for the peculiar nucleosynthesis paths that can be enabled.\\
A 4$\pi$-3D hydrodynamic simulation of an oxygen-neon shell merger has been run by \cite{Yadav_2020}. Differently from the previous studies, in this work the shells start as convective and independent, and dynamically merge during the simulation. \cite{Yadav_2020} found very strong differences between the 1D and 3D simulations, starting from the very large convective velocities compared to the 1D mixing-length-theory predictions. However, their merging takes place only within about 200 seconds before the onset of core collapse, therefore the merging time-scale is very limited and the system has only time to present an episodic burning of ingested neon. Additionally, the preliminary contraction of the shells is the primary driving mechanism for the merging \citep[see also][]{10.1093/mnras/stx2470}. This is not always the case for shell mergers in massive stars: in particular, several 1D models predict shell merging events that occur many convective turnovers before the stellar collapse, with enough time for a new equilibrium structure to be established, with important implications for the final structure of the star. The merging of independent convective shells long before the stellar collapse has never been investigated in the literature with multi-D models.\\
In this paper, we present the results from a set of 3D hydrodynamic models simulating a shell merging event predicted by a 1D model, 5 hours before the predicted collapse of the star. By analysing both the dynamics and the nucleosynthesis of this environment, we are able to shed light on these poorly explored shell-merging events, drawing conclusions of interest for stellar structure and chemical evolution theory. We organize the paper as follows: in Section 2, we introduce the model setup and initial conditions used to run the simulations. In Section 3, we present the results divided into the analysis of the dynamics and of the nucleosynthesis of the simulations. Finally, in Section 4 we discuss the results and draw conclusions.

\section{Methods}
\begin{figure*}
\centering
\footnotesize
\includegraphics[trim={0.33cm -0.25cm 1.25cm 0cm},clip,width=0.56\textwidth]{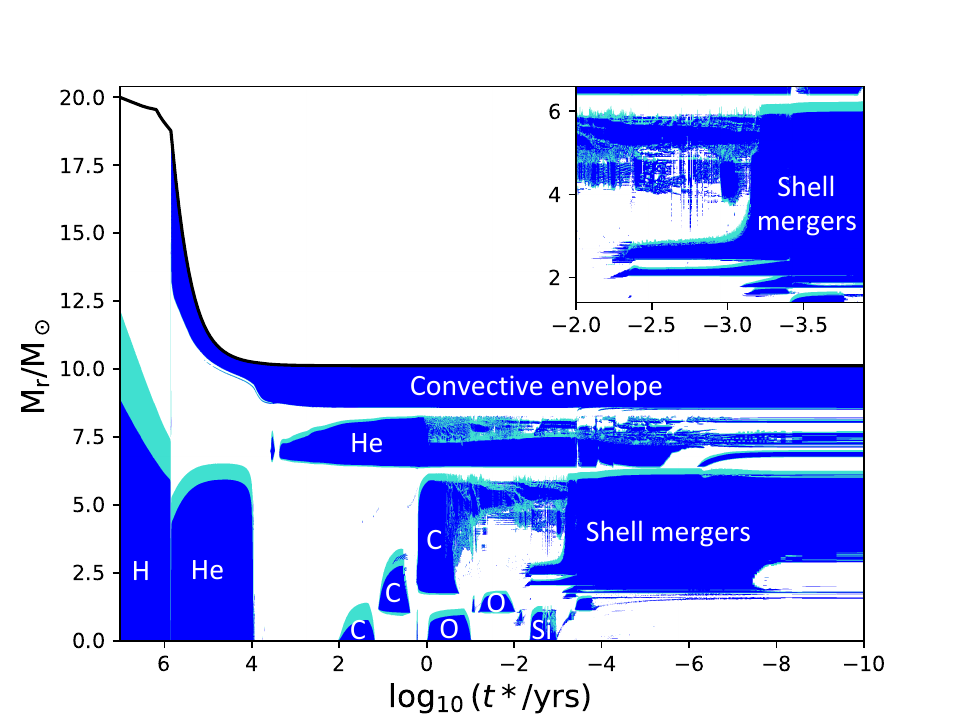}
\includegraphics[trim={0.25cm 0cm 0cm 0cm},clip,width=0.435\textwidth]{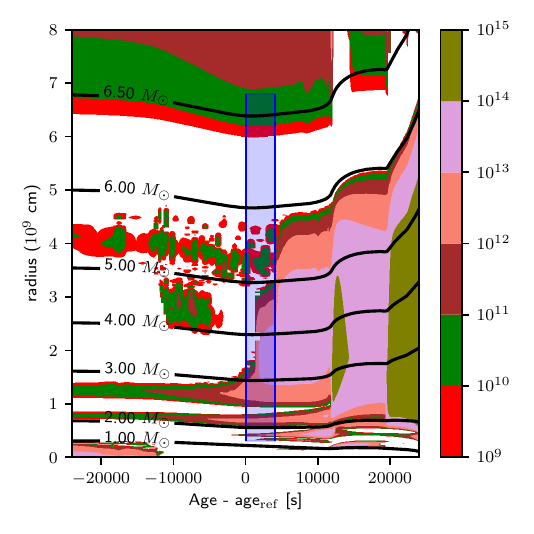}\put(-336,33){\color{red}\Huge\boldmath$\uparrow$}
\caption{{\it Left}: Structure evolution diagram of the 20 M$_\odot$ 1D \texttt{MESA} model as a function of the time left until the predicted collapse of the star (in years, log scale). Convective zones are drawn in blue, CBM zones in green. The red arrow indicates the shell-merging event, of which a zoom-in is shown in the top right corner. {\it Right:} Zoom-in on the shell-merging event, as a function of time in seconds from the start of the 3D simulations. In colour scale, the squared convective velocity (cm s$^{-1}$). The black lines are isomass contours. The vertical blue bars represent the radial and time extent of the 3D simulations.}\label{fig:kip}
\end{figure*}

\subsection{The 1D stellar model and initial conditions}
The initial conditions for the 3D simulations presented in this paper have been assumed from the same 1D stellar evolution model used in \cite{2023Rizzuti}; it is worth summarizing here its most important aspects. This is a \texttt{MESA} \citep{Paxton2011, Paxton2013, Paxton2018, Paxton2019} stellar evolution model of a 20 M$_\odot$ star at solar metallicity ($Z=0.014$, \citealt{Asplund}). Mass loss rates for O-type stars are assumed from \cite{2000A&A...362..295V, 2001V}; if the star enters the Wolf-Rayet stage, i.e.\ when the surface hydrogen mass fraction drops below 0.4, the mass loss rate switches to \cite{2000A&A...360..227N}; if $T_\text{eff} < 10^4 \text{ K}$, the empirical mass loss rate from \cite{1988A&AS...72..259D} is used. The mixing-length theory \citep[MLT,][]{1958ZA.....46..108B} of convection is employed (using the “Henyey” and “MLT++” options), with an efficiency of $\alpha_\text{MLT}=1.67$ \citep{2018arXiv181004659A}. The convective boundaries are defined by the Schwarzschild criterion, so no semi-convective mixing is required. For convective boundary mixing, the model uses the exponential decaying diffusive prescription of \cite{1996A&A...313..497F} and \cite{2000A&A...360..952H}, with $f_\text{ov} = 0.05$ for the top of convective cores and shells, and $f_\text{ov} = 0.01$ for the bottom of convective shells (with $f_0 = f$ in both cases). This implementation of convective boundary mixing is the real novelty of this model, as presented in \cite{2023Rizzuti}. The values chosen for $f_\text{ov}$ are larger than what currently used in the grids of stellar models, e.g.\ $\alpha_\text{ov}=0.1$ in \citealt{2012Ek} and $\alpha_\text{ov}=0.335$ in \citealt{2011B}, considering that $f_\text{ov}\sim\alpha_\text{ov}/10$ \citep[see][]{2021Scott}. Choosing $f_\text{ov}= 0.05$ here is motivated by the study of \cite{2021Scott}, which predicts values for $f_\text{ov}$ of at least 0.05 for 20 M$_\odot$ stars in order to reproduce the observed width of the main sequence in the spectroscopic Hertzsprung-Russell diagram \citep{2014Ca}. For the bottom boundary, $f_\text{ov}= 0.01$ is based on 3D hydrodynamic results \citep{2019Cristini, 2022Rizzuti}, that find a weaker convective boundary mixing (CBM) at the bottom boundary due to it being stiffer. These choices represent what we call the `321D' approach, where results from hydrodynamic simulations are used to improve the prescriptions assumed in the 1D models. In support of this, \cite{2021Scott} showed that CBM increases with the initial stellar mass since more massive stars are much more luminous ($L\sim M^3$ between 1 and 20 M$_\odot$); for this reason, our choice of $f_\text{ov}= 0.05$ for 20 M$_\odot$ is consistent with the values around $f_\text{ov} = 0.02\text{ - }0.04$ inferred from asteroseismology for less massive stars \citep[see][]{2020Bow}. This new way of modelling CBM, which is derived from 3D hydrodynamic simulations but is also consistent with observations, is what makes this model novel. The amount of CBM to be included in stellar models is extremely important for the occurrence of shell merging events, because it is entrainment that erodes the radiative regions that separate the convective ones, and makes it possible for the merging to occur. This has been shown for example in \cite{10.1093/mnras/sty3415}, where the model with the largest CBM is also the one that shows the occurrence of C-Ne shell merging, though more studies are needed to confirm these trends. \\
In Fig.~\ref{fig:kip}, we show the structure evolution diagram of the model, also represented in \cite{2023Rizzuti}. We focus our attention here on the shell-merging event, which takes place between $10^{-2}\text{ - }10^{-4}$ years before the predicted collapse of the star; we present a zoom-in of this region in the right panel of Fig.~\ref{fig:kip}. The merging occurs about 5 hours before the collapse of the star, which is much longer than the convective turnover time-scale, typically up to a couple hundred seconds; this means that the core contraction is not a driving mechanism of the merging \citep[as in e.g.][]{Yadav_2020}. From these plots, it is possible to see the presence of three distinct convective regions at the start of the 3D simulations, and they are all predicted to merge after about $12\ 000\text{ s}$. A fourth convective shell forms below the others after the 3D simulations have started, but it does not join the merging and halts around 10 000 s (see also Fig.~\ref{fig:tke_shell} in the next section). The first question that the hydrodynamic simulations shall answer is whether a merging also takes place in 3D within the simulated time range or not. It is still not completely clear whether shell merging is a just a numerical effect of the 1D models, or this phenomenon is also expected to occur in real stars: hydrodynamic simulations of multiple convective shells will be able to shed more light on this point. Finally, Figure \ref{fig:kip} includes some isomass contours to show that some expansion of the layers occurs during the merging event: to account for that in the 3D simulations, a large radial extent has been selected for the 3D domain.

\begin{figure*}
\centering
\footnotesize
\includegraphics[trim={4cm 4cm 4cm 4cm},clip,width=0.6\textwidth]{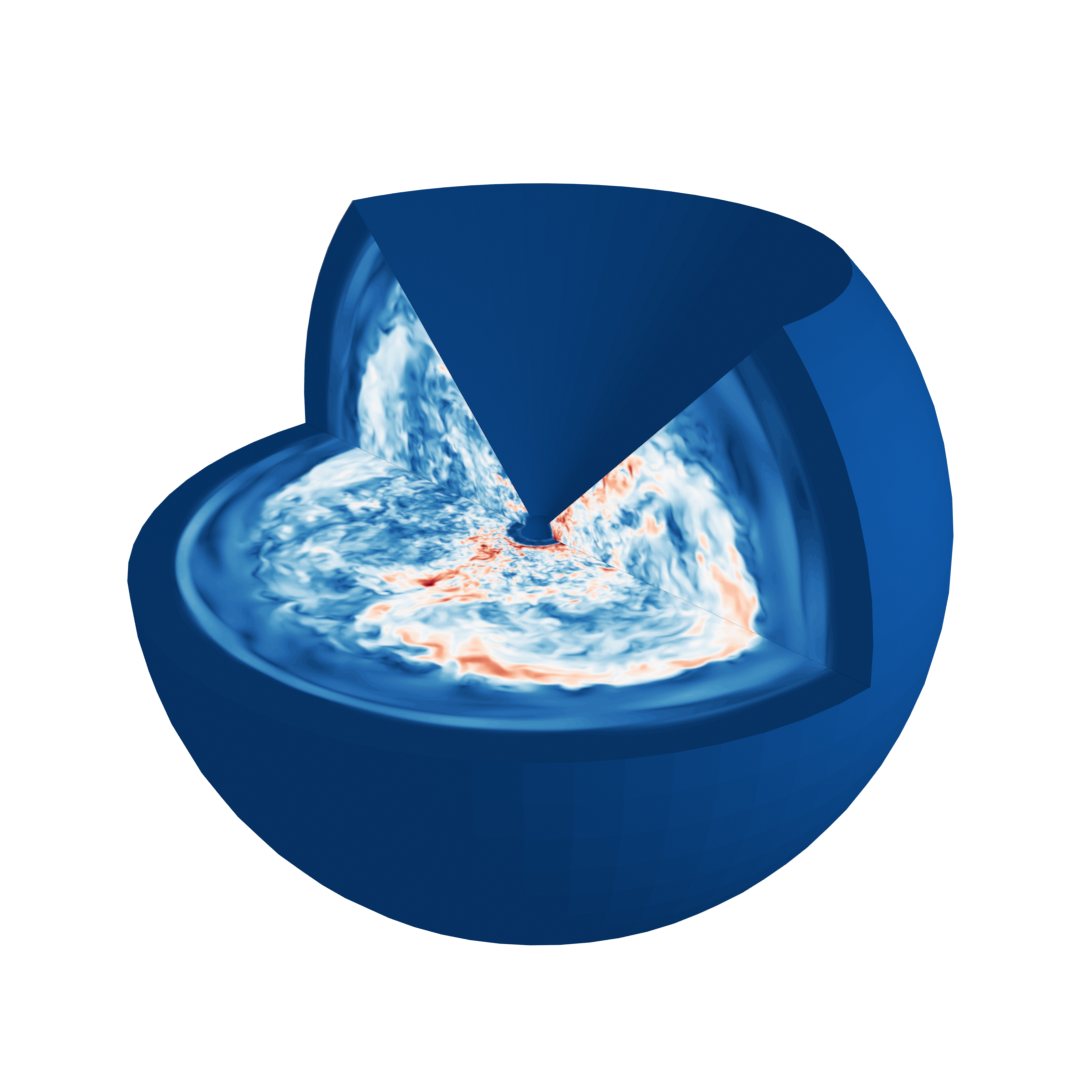}
\includegraphics[trim={39cm 0.5cm 0cm 0cm},clip,width=0.13\textwidth]{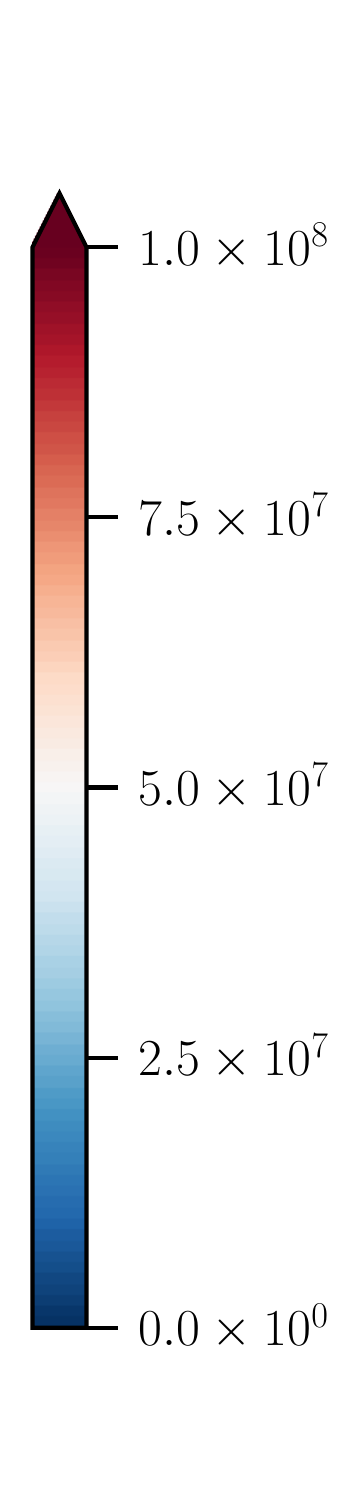}
\caption{3D rendering of model \texttt{a360n1024} showing the simulation through a cross section along both the equatorial and the longitudinal planes. The frame has been taken at 1500 seconds, and the fluid speed is in colour scale (cm s$^{-1}$). A complete video of the time evolution for this rendering is available online as Supplementary material.}\label{fig:model}
\end{figure*}
\subsection{The 3D model domain and configuration}
The 3D hydrodynamic simulations presented in this study have been produced employing the \texttt{PROMPI} hydrodynamic code \citep{Meakin_2007}. Its established efficiency and adaptability in performing multi-dimensional simulations, together with its long history of simulating stellar environments \citep{Arnett_2009,2017Cristini,2019Cristini,Mocak2018,2022Rizzuti,2023Rizzuti}, make \texttt{PROMPI} the ideal tool for conducting our study. In particular, active tracers for composition and nuclear network as described below are key aspects of the code. Additionally, the code comparison study of \cite{Andrassy2022} showed that \texttt{PROMPI} is fully consistent with other hydrodynamic codes commonly employed for stellar studies.\\
We list here the complete set of combustive Euler equations solved by \texttt{PROMPI}, as they are described in detail in \cite{Meakin_2007}, in state-vector form with \textbf{\textit{Q}} the state vector, $\mathbf{\Phi}$ the flux vector, and \textbf{\textit{S}} the source vector:
\begin{equation}
\begin{aligned}
&\frac{\partial{\textbf{\textit{Q}}}}{\partial{t}}+\nabla\cdot\mathbf{\Phi} = \textbf{\textit{S}}\\
&\textbf{\textit{Q}} = \begin{cases}
        \rho \\
        \rho \textbf{\textit{v}}\\
        \rho E\\
        \rho X_i
    \end{cases}
\mathbf{\Phi} = \begin{cases}
\rho \textbf{\textit{v}}\\
\rho \textbf{\textit{v}}\cdot\textbf{\textit{v}} + p\\
(\rho E + p)\textbf{\textit{v}}\\
\rho X_i \textbf{\textit{v}}
\end{cases}
\textbf{\textit{S}} = \begin{cases}
        0\\
\rho \textbf{\textit{g}}\\
\rho \textbf{\textit{v}} \cdot \textbf{\textit{g}} + \rho \epsilon\\
R_i\\
    \end{cases}
\end{aligned}
\end{equation}
where $\rho, p, \textbf{\textit{v}}, \textbf{\textit{g}},$ and $T$ the density, pressure, velocity, gravity, and temperature, respectively. $E$ is the total specific energy, and the energy source term $\epsilon$ is due to nuclear reactions and neutrino cooling. $R_i$ is the time rate of change of species $X_i$ due to nuclear reactions.
\\When remapping the initial conditions from the 1D into the 3D model, one needs to be careful to ensure accuracy and consistency. As it is standard procedure in \texttt{PROMPI}, the hydrostatic equilibrium of the stratification has been recomputed by obtaining new density and temperature values from pressure and entropy through the equation of state, checking that it does not deviate significantly from the original values. \texttt{PROMPI} employs the `Timmes' equation of state \citep{Timmes_1999}.\\
As we showed in Fig.~\ref{fig:kip}, left, for the 3D simulations we selected a radial extent of $0.3<r<6.8\times 10^9$ cm, including enough space above the convective shells to account for the upward expansion of the layers. This domain includes in the outermost part some He-rich convective layers, which are excluded from the simulations by implementing a velocity-damping region at $r>6\times 10^9$ cm, also used to dissipate the gravity waves produced by the convective boundaries; the damping function used by \texttt{PROMPI} is the one described by \cite{2017Cristini}. Finally, convection is triggered by seed perturbations added to density and temperature between 6 - $7 \times 10^8$ cm (Ne-burning shell) and between 9.5 - $12 \times 10^8$ cm (C-burning shell), as described by \cite{Meakin_2007}.
\\In this study, we present two different simulations of the shell merging event, both started from the same initial conditions but with different geometry. Both simulations have spherical geometry and a radial extent of $0.3<r<6.8\times 10^9$ cm, but the angular range covered by $\theta$ and $\phi$ is different. One setup is a 3D wedge with an angular size of $60^{\circ}$ in both $\theta$ and $\phi$, and number of cells $768\times 256^2$ in $r, \theta$ and $\phi$, respectively; we code-name this model \texttt{a60n256} after its angular and grid size. The second model has instead an angular size of $90^{\circ}$ in $\theta$ and $360^{\circ}$ in $\phi$, and number of cells $512\times 256\times 1024$ in $r, \theta$ and $\phi$, respectively; we code-name this \texttt{a360n1024}. To give a visual representation of this setup, we show in Fig.~\ref{fig:model} a 3D rendering of the domain of model \texttt{a360n1024}, including a cross section that shows an equatorial and a longitudinal view of the simulation.\\
The reason for running these two different simulations is to study the evolution of the same initial conditions in the two geometries, in order to test convergence of results and evolutionary divergences. In particular, \texttt{a60n256} has a higher local resolution but more limited spatial extent, while \texttt{a360n1024} is closer to a full sphere, covering over 70 per cent of the spherical surface, at the cost of a slightly lower local resolution. The reason why the \texttt{PROMPI} code cannot perform full 4$\pi$ simulations is the presence in the spherical grid of singularities at the centre and along the polar axis, therefore artefacts are produced by the code the closer the domain approaches these points in space. However, we underline here the importance of going beyond the box-in-a-star setup and towards fully spherical simulations, in order to correctly reproduce the fluid motions especially in case of large convective regions.\\
In both models, periodic boundary conditions have been implemented in $\phi$, but for $\theta$ we chose reflective boundary conditions instead, due to the close proximity to the polar axis, where periodic conditions are no longer realistic and can create an excess of kinetic energy, therefore it is more physical to assume that the flow cannot cross the axis \citep[see e.g.][]{2020LRCA....6....3M}. \\
Finally, one of the key strengths of the \texttt{PROMPI} code is 
that it explicitly models the evolution of chemical species, used in this study as active tracers. For the simulations presented here, a 12-isotope network has been employed to reproduce nuclear reactions and generate the energy that drives convection. This network, which is the same used in \cite{2023Rizzuti}, includes n, p, $^4$He, $^{12}$C, $^{16}$O, $^{20}$Ne, $^{23}$Na, $^{24}$Mg, $^{28}$Si, $^{31}$P, $^{32}$S, and $^{56}$Ni, and it employs the most recent nuclear rates from the \texttt{JINA REACLIB} data base \citep{Cyburt_2010}. This list of isotopes is particularly appropriate for reproducing the shell-merging environment, because it can follow all the energy-generating reactions that comprise carbon, neon and oxygen burning, which are expected to take place in these convective shells. The shell-merging environment is dynamical enough that boosting the driving luminosity (used in some simulations e.g.\ \citealp{2017Cristini,2019Cristini}) is not required here, ensuring that no artefacts arise from the changes in the energy generation.

\section{Results}
\subsection{Dynamics of the shell merging}
\begin{figure*}
\centering
\footnotesize
\includegraphics[trim={0cm 0cm 0cm 0cm},width=0.75\textwidth]{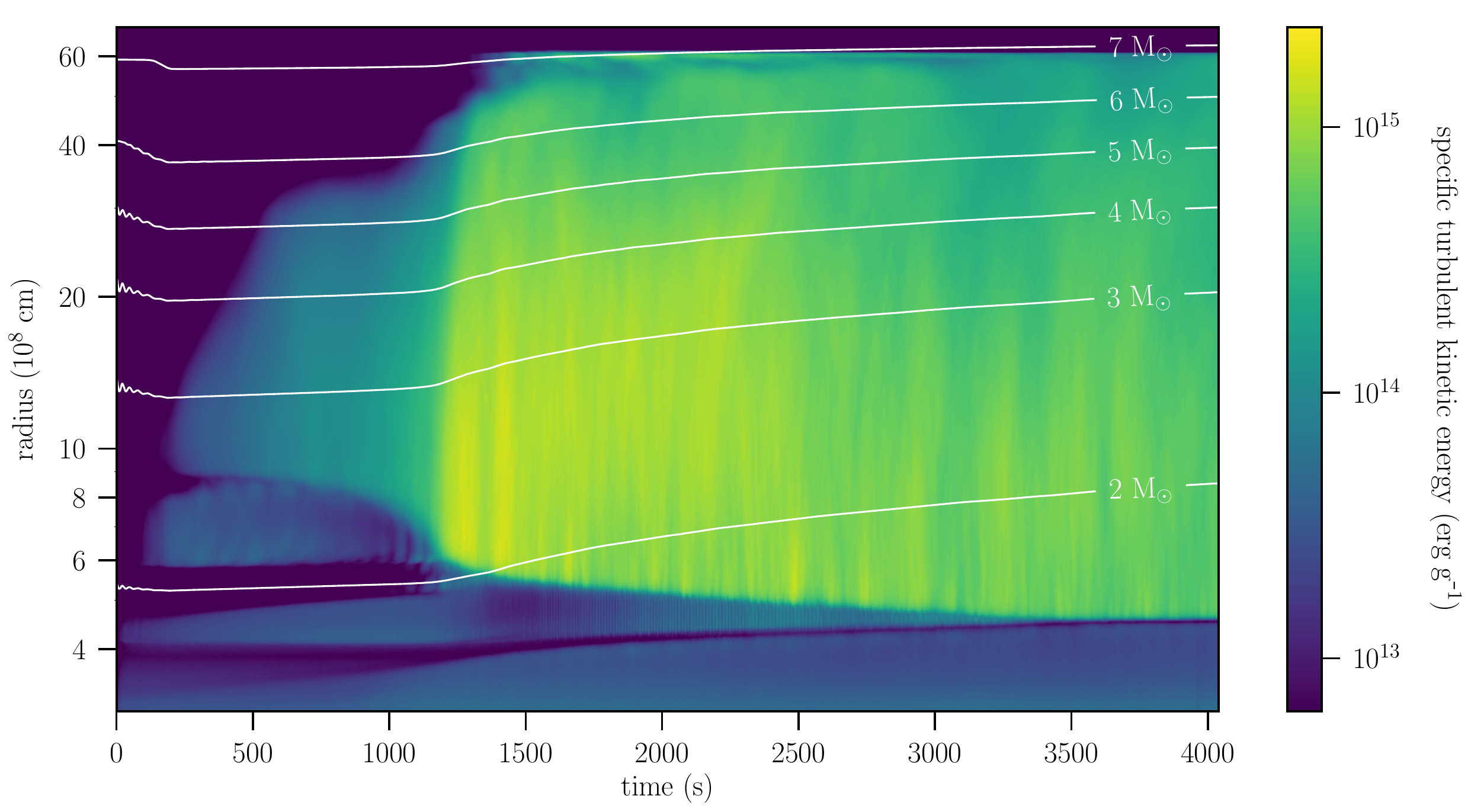}
\put(-340,178){\color{white}\huge\texttt{a360n1024}}
\put(-315,116){\color{white}\huge\texttt{C}}
\put(-328,70){\color{white}\huge\texttt{Ne}}
\put(-332,43){\color{white}\huge\texttt{O}}\\
\includegraphics[trim={0cm 0cm 0cm 0cm},width=0.75\textwidth]{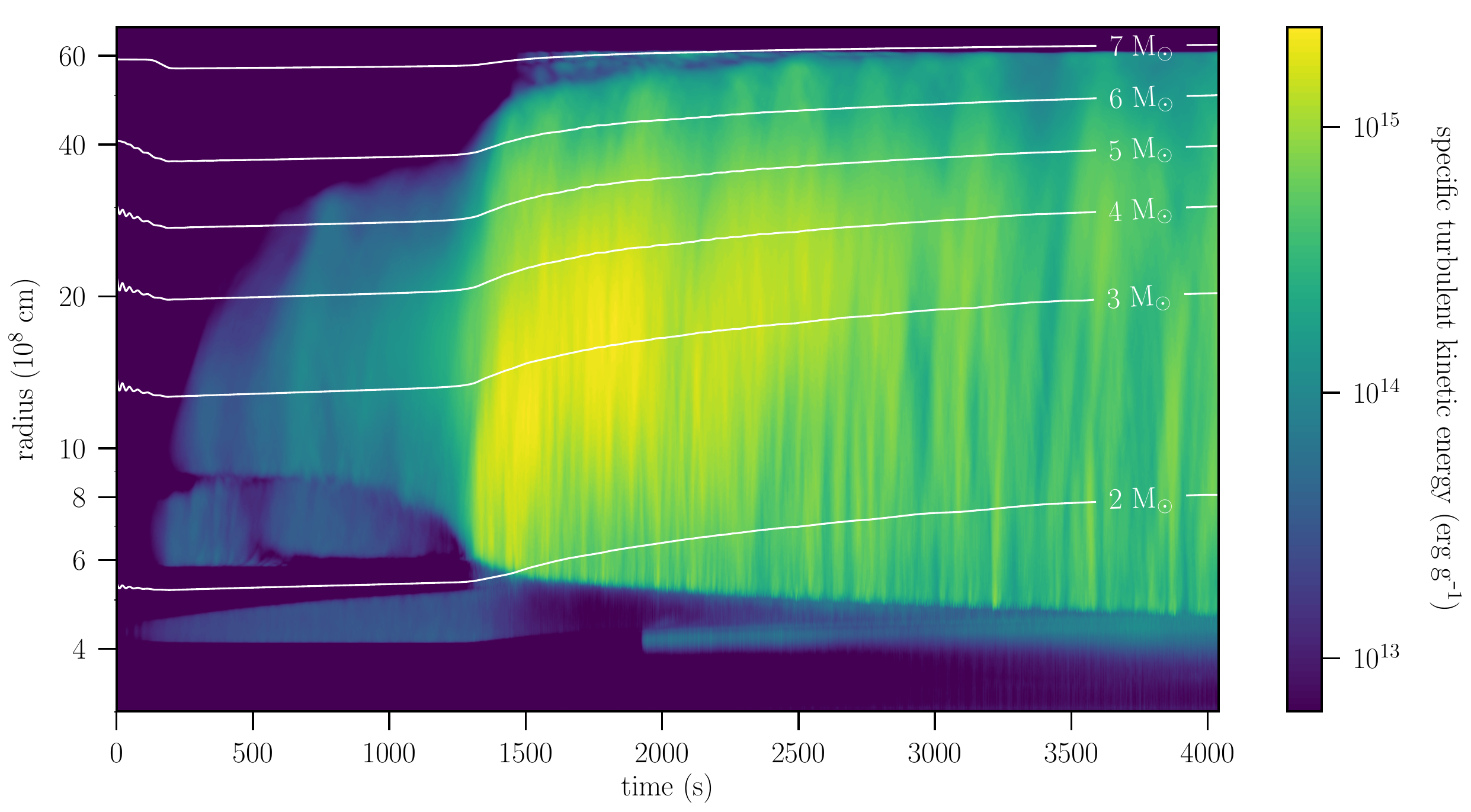}
\put(-332,178){\color{white}\huge\texttt{a60n256}}
\put(-315,116){\color{white}\huge\texttt{C}}
\put(-328,70){\color{white}\huge\texttt{Ne}}
\put(-332,43){\color{white}\huge\texttt{O}}\\ 
\includegraphics[trim={0cm 0cm 0cm 0cm},width=0.75\textwidth]{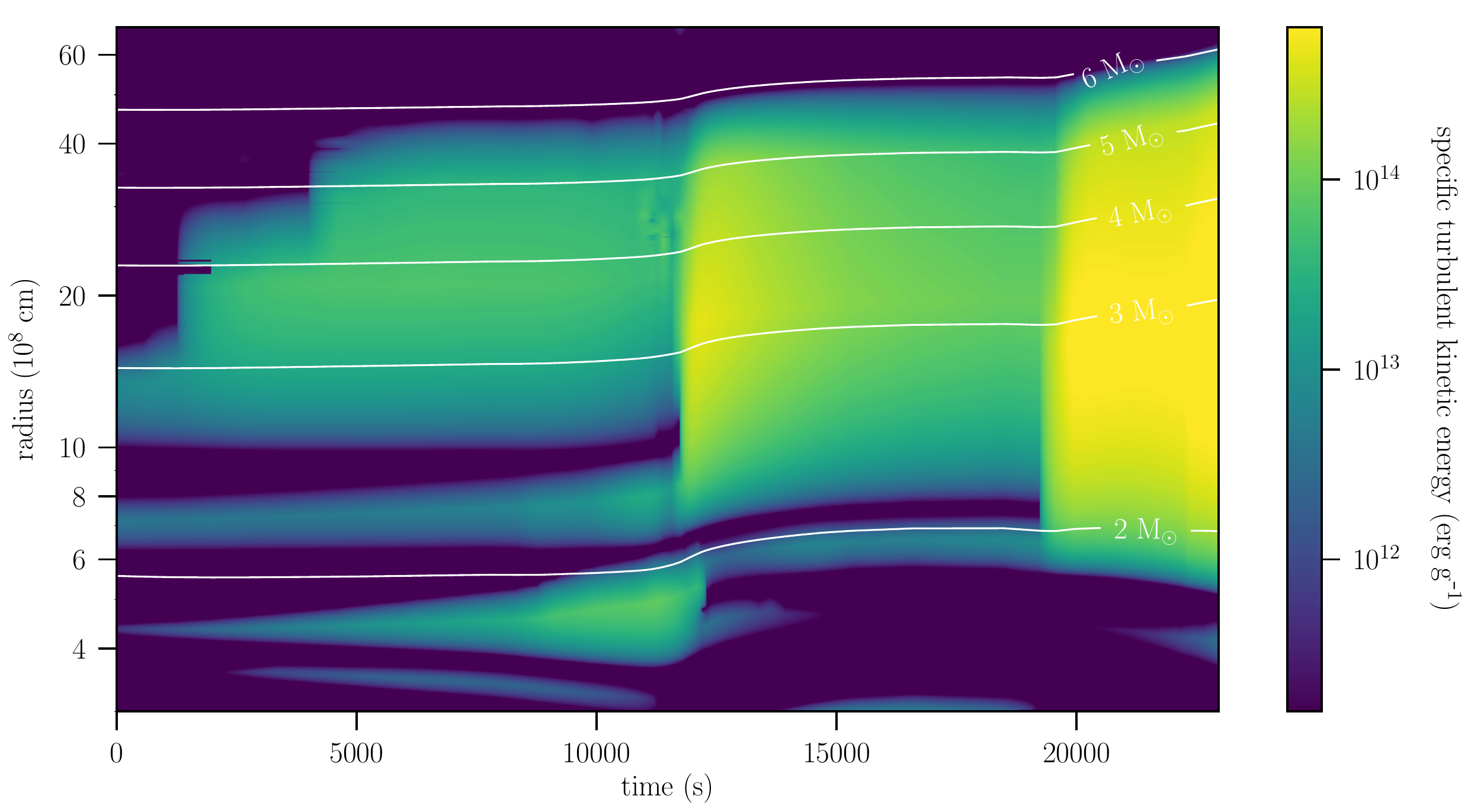}
\put(-332,186){\color{white}\huge\texttt{1D MESA}}
\put(-315,124){\color{white}\huge\texttt{C}}
\put(-328,70){\color{white}\huge\texttt{Ne}}
\put(-332,43){\color{white}\huge\texttt{O}}
\caption{Time evolution of the angularly averaged kinetic energy in colour scale for 3D simulations \texttt{a360n1024} (top panel), \texttt{a60n256} (centre panel), and the 1D \texttt{MESA} model for comparison (bottom panel). Overlaid in white are the isomass contours. The log scale applied to the radius on the y-axis provides a comprehensive view of all the convective shells.}\label{fig:tke_shell}
\end{figure*}
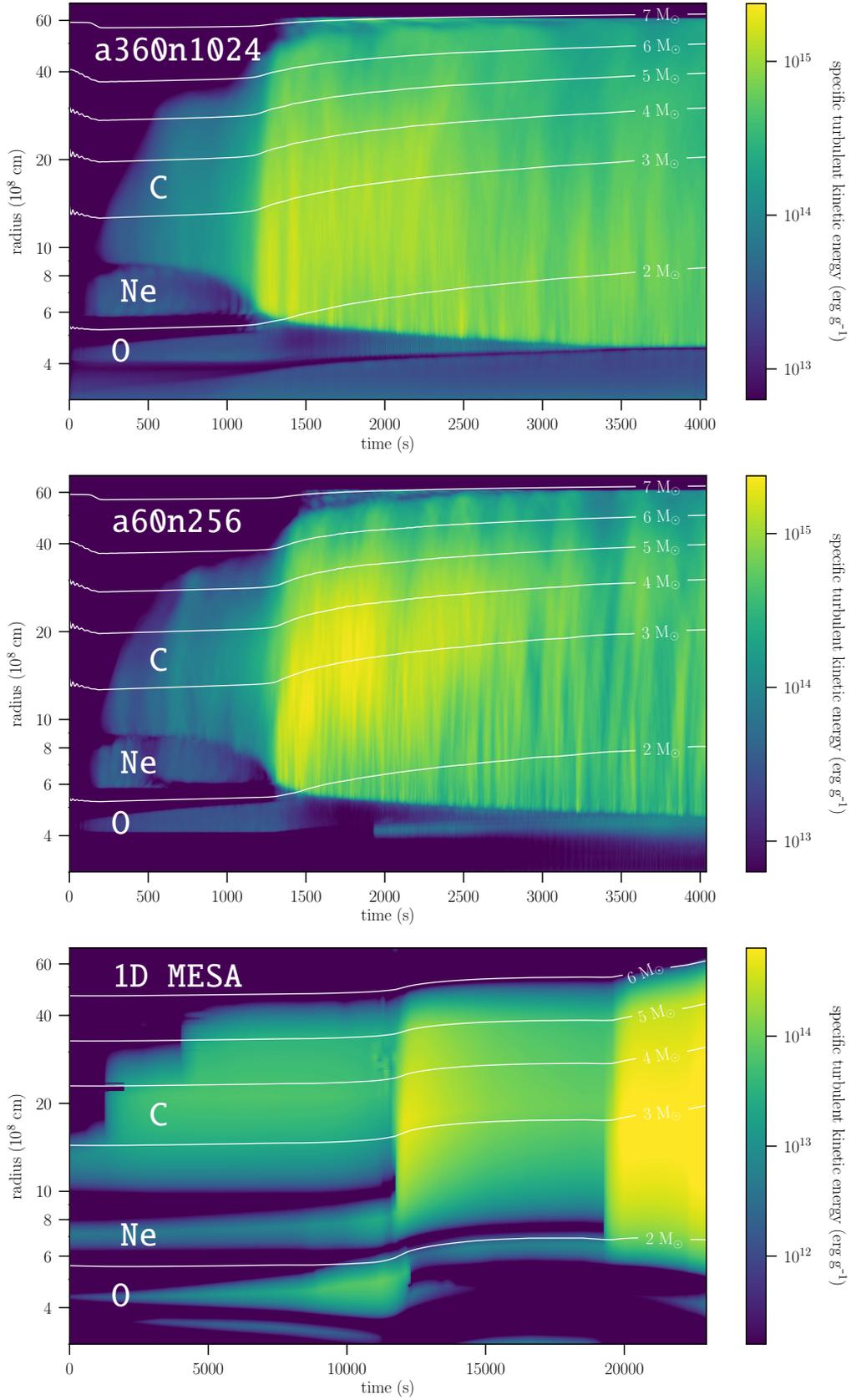

\begin{figure*}
\centering
\footnotesize
\includegraphics[trim={0cm 0cm 0cm 0cm},width=\textwidth]{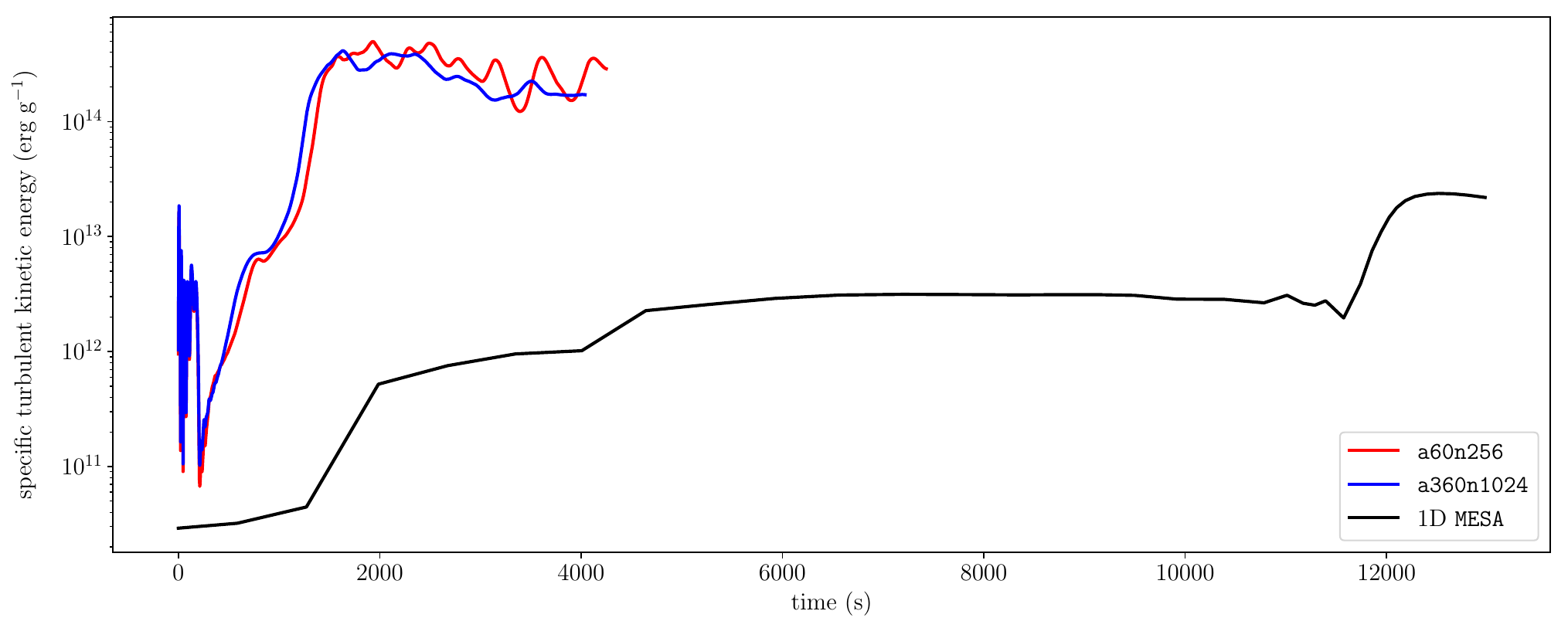}
\caption{Specific turbulent kinetic energy integrated across the entire domain for simulations \texttt{a360n1024} and \texttt{a60n256}, compared to the same quantity in the 1D \texttt{MESA} model, integrated over the same domain.}\label{fig:tke_shell_time}
\end{figure*}

\begin{table}
\centering
\footnotesize
\caption{Properties of the hydrodynamic simulations presented in this study: model code name; polar angular extent $\Delta\theta$; azimuthal angular extent $\Delta\phi$; number of radial cells $N_{r}$; number of polar angle cells $N_{\theta}$; number of azimuthal angle cells $N_{\phi}$; end time of the simulation $t_\text{end}$; cost required to run the simulation in CPU core-hours.}\label{tab:1}
\begin{tabular}{cccccccc}
\hline\\[-0.7em]
name&$\Delta\theta$&$\Delta\phi$&$N_{r}$&$N_{\theta}$&$N_{\phi}$&$t_\text{end}$&cost\\
&&&&&&(s)&($10^6$ hr)\\ \\[-0.7em]\hline\\[-0.7em]
\texttt{a60n256}&$60^{\circ}$&$60^{\circ}$&768&256&256&4250&1.31\\
\texttt{a360n1024}&$90^{\circ}$&$360^{\circ}$&512&256&1024&4039&3.34\\
\\[-0.7em]\hline
\end{tabular}
\end{table}
We present in Table \ref{tab:1} the main properties of the two hydrodynamic simulations included in this study. The two simulations are started from the same initial conditions and run for the same time-scale, but with two different geometries. As we show below, the evolution of the two simulations is very similar, so when the analysis is not focused on the differences arising from the geometry, we prefer to show results only from model \texttt{a360n1024}, which is closer to a full sphere. As an example, we have shown in Fig.~\ref{fig:model} a representation of model \texttt{a360n1024} containing two cross-sections, one across the equatorial plane and the other across the longitudinal plane; these are shown in more detail in Appendix \ref{append:A}, Fig.~\ref{fig:360frame}. It is clear the effect of the geometry on the fluid motions: from the equatorial plane, we see large-scale structures that can form thanks to the large radial extent and the 360$^{\circ}$ range spanned by $\phi$. In the vertical plane instead, large-scale eddies take up the entire domain, mainly due to the reflective boundary conditions assumed in $\theta$, that encourage the formation of one large eddy. Additional tests show that implementing periodic boundary conditions in $\theta$ encourages instead the formation of two large eddies within the same convective region. This finding is in line with the general physical expectations for similar environments (non-rotating stars with large radial extent in deep interiors), as e.g.\ in the convective core study of \cite{herwig20233d}. This is not necessarily the case in other convective environments, for example in envelope convection, even when the radial extent is large \citep[see][]{refId0,refId1}.
\\We shall now study the evolution of the multiple convective shells in the 3D simulations. To have a visual representation similar to the stellar evolution diagram showed in Fig.~\ref{fig:kip} for the 1D model, we present in Fig.~\ref{fig:tke_shell} the time evolution of the angularly averaged kinetic energy (in colour scale) for the two hydrodynamic models \texttt{a360n1024} (top panel) and \texttt{a60n256} (centre panel), compared to the same diagram for the 1D \texttt{MESA} model (bottom panel). To have a comprehensive view on the different convective shells and track their evolution, we applied a log scale to the stellar radius. The main event in all simulations is the merging of the carbon- and the neon-burning shells, generating a large increase in the kinetic energy due to the burning of the freshly engulfed material. 
The carbon and neon shells merge both in the 3D and 1D simulations, there are nevertheless significant differences between the two types of simulations. 
One major difference is that only in 3D the bottom of the carbon shell migrates downwards into the neon-shell until the C- and Ne-rich material from the C-burning shell reaches the bottom of the Ne-burning shell, triggering the rapid burning of the fresh fuel and nuclear energy release, visible as a sharp increase in kinetic energy in Fig.~\ref{fig:tke_shell}. This scenario underlines the necessary role of entrainment processes in the advanced phases of massive stars, as it has been studied by works such as \cite{Meakin_2007}, \cite{Viallet}, \cite{2023Rizzuti}.
\\Additionally, in the 1D model CBM is included using the exponentially decaying diffusion from \cite{1996A&A...313..497F} and \cite{2000A&A...360..952H}. This leads to a slower growth of the neon-shell in the 1D simulation and thus a later merging in 1D compared to 3D (see Fig.~\ref{fig:tke_shell}), around 12 000 s for the former and only 1200 s for the latter, from the start of the 3D simulations. This is illustrated also in Fig.~\ref{fig:tke_shell_time}, where we show the time evolution of the integrated kinetic energy for the 3D versus 1D simulations: in addition to the different time-scale before the merging (recognisable by the sharp increase in kinetic energy), the 3D simulations reach a total kinetic energy that is around one order of magnitude larger than in the 1D, also visible by the colour scale in Fig.~\ref{fig:tke_shell}. It is not easy to immediately understand this difference, given the complexity of this environment and the interplay of different effects (entrainment, nuclear burning, convective velocities) across the multiple burning shells. First, we must consider that the stratification assumed from the 1D model does not result in equilibrium once in the hydrodynamic model, giving way to a slight readjustment of the structure during the initial transient phase. This can be traced back, among other things, to the fact that the 1D model assumes hydrostatic equilibrium, which is not necessarily accurate for this late dynamical phases. The deviations are smaller in the inner regions and grow larger going outwards, due to the recomputation. Indicatively, the difference in density and temperature is below 10 and 4 per cent, respectively, in the inner regions, and it reaches a maximum around 25 and 8 per cent in the outer regions. Note that the outer regions do not play any significant role in our investigation, so these large deviations do not change the results presented.
The slight contraction during the initial transient (visible also from Fig.~\ref{fig:tke_shell}) results in 3D peak temperatures around 5-10 per cent higher, depending on the shell; this affects also the nuclear and kinetic energy, and therefore the time-scale of evolution. However, given the multiple burning reactions and dependencies, it is difficult to assign exact numbers to these estimates.
\begin{figure*}
\centering
\footnotesize
\includegraphics[trim={0cm 0cm 0.15cm 0cm},width=\textwidth]{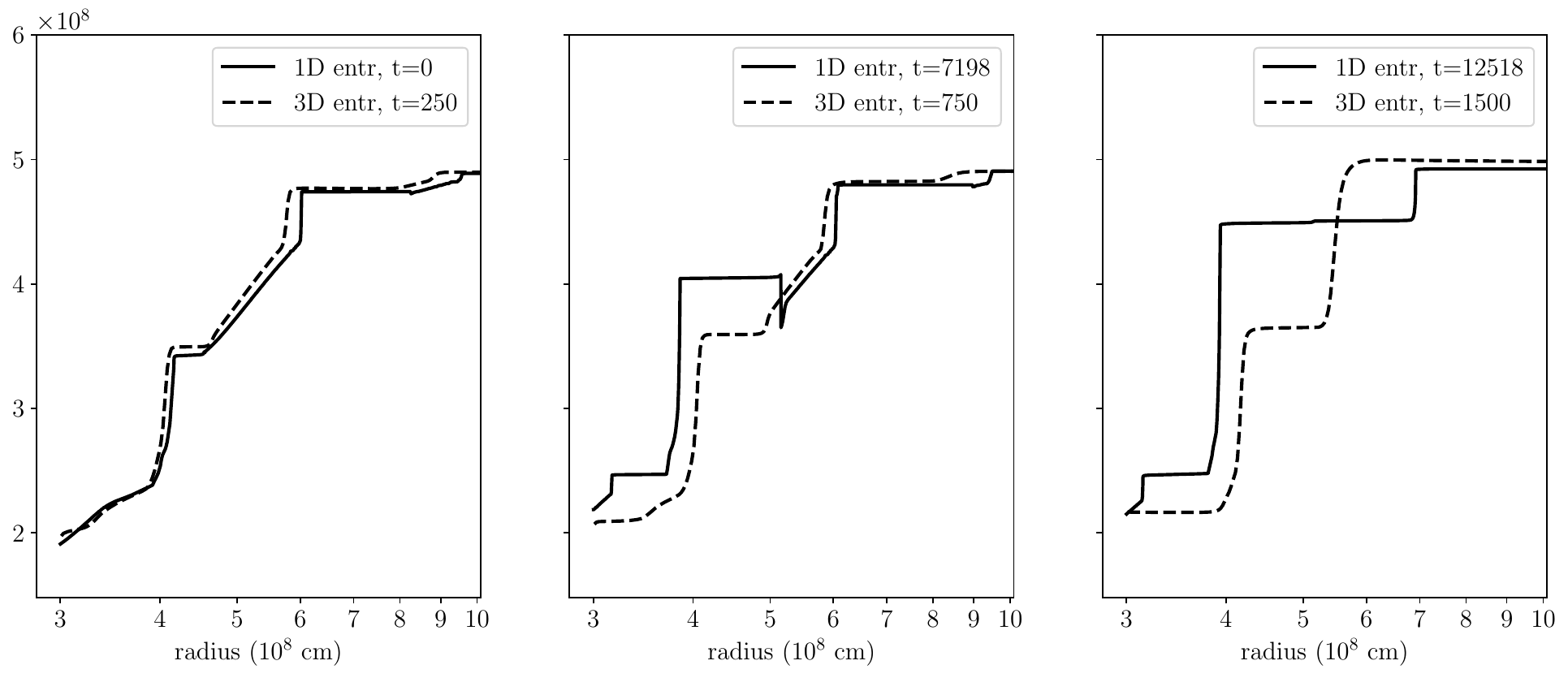}
\caption{Angularly averaged radial profiles of entropy (erg K$^{-1}$), comparison between 1D \texttt{MESA} model (solid) and 3D model \texttt{a360n1024} (dashed), at different times throughout the simulations: initial conditions (left), before (centre) and after (right) the shell merging event.}\label{fig:entropy}
\end{figure*}
\\Finally, we shall now focus our attention on the innermost convective shell in the domain, the oxygen-burning shell: its behaviour is different in the three models presented in Fig.~\ref{fig:tke_shell}. What all the simulations have in common is that the O-shell evolution is suddenly halted at the time of the shell merging; while in 3D this is due to the downward migration of the shell merger, in 1D this is more likely the result of the expansion of the layers due to the formation of a Si-burning shell underneath, producing the same result. After the merging, the fate of the O-shell is different across the models: in 1D, a shallow O-burning layer survives and later merges with the shell above, producing a second important merging around 20 000 s. This second merging is absent in the 3D simulations, where instead the O-shell slowly turns off and is entrained by the shell merger above. To better understand what is happening here, we plot in Fig.~\ref{fig:entropy} the evolution of the different entropy profiles in the 1D versus the 3D simulations. Here and in other parts of this paper, we compare the 1D and 3D simulations at different time-steps, given the different time-scale of evolution between the two; the selected time-steps are listed in Table \ref{tab:1.5}.
\begin{table}
    \centering
\footnotesize
\caption{Selected time-steps for comparing the 1D and 3D simulations.}\label{tab:1.5}
    \begin{tabular}{l|ccc}
         &initial&before&after\\
         &conditions&merging&merging\\ \\[-0.7em]
         \hline \\[-0.7em]
         1D \texttt{MESA}&0 s&7198 s&12518 s\\
         3D \texttt{PROMPI}&250 s&750 s&1500 s
    \end{tabular}
\end{table}
From the figure, we can highlight the differences between the simulations: starting from a similar configuration (Fig.~\ref{fig:entropy}, left), the plateau corresponding to the O-shell between 4 - $5\times 10^8$ cm grows both in radius and in magnitude over time, but much larger and higher in the 1D model than in 3D. This can be seen as an effect of the lack of entropy mixing in the \texttt{MESA} code, as we already highlighted in \cite{2023Rizzuti}, producing the artefacts visible in Fig.~\ref{fig:entropy}, centre. The net result is that the O-shell entropy plateau in 1D can easily overcome the barrier that separates it from the shell merger at $r>7\times 10^8$ cm (Fig.~\ref{fig:entropy}, right) and produce a second merging, as visible in Fig.~\ref{fig:tke_shell}, while in 3D the weaker evolution of the entropy plateau preserves a strong entropy barrier that prevents the second merging. At later times in 3D, the oxygen burning definitely stops and the O-shell is eventually entrained by the shell merger (see Fig.~\ref{fig:tke_shell}).
\\It is worth mentioning here also the presence of a second, weaker convective shell below the oxygen-burning one in model \texttt{a360n1024}, which is not present instead in model \texttt{a60n256} (see Fig.~\ref{fig:tke_shell}). This might be the result of the convectively unstable stratification that generates a second oxygen-burning shell there in the 1D model, but this is more likely an effect of the geometry of the simulation, which going towards the polar singularity in the spherical grid may be generating these artefacts. Anyway, we do not treat this shell as physical, and it is not included in the analysis. 

\begin{figure*}
\centering
\footnotesize
\includegraphics[trim={0cm 0cm 0cm 0cm},width=\textwidth]{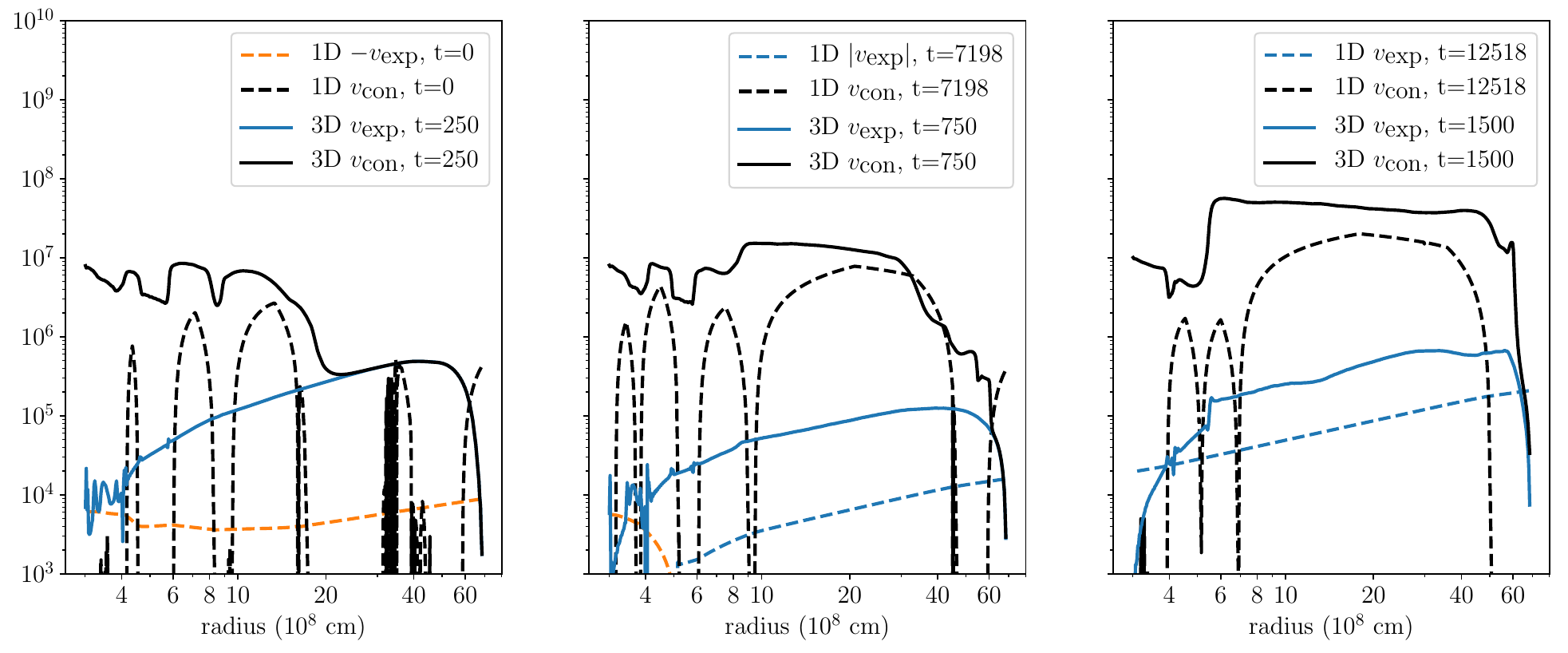}
\caption{Angularly averaged radial profiles of the expansion and convective velocities in cm s$^{-1}$ (see text for definitions), comparison between 1D \texttt{MESA} model (dashed) and 3D model \texttt{a360n1024} (solid), at different times throughout the simulations: initial conditions (left), before (centre) and after (right) the shell merging event.}\label{fig:velocities}
\end{figure*}
\subsection{The velocity field}
\begin{figure}
\centering
\footnotesize
\includegraphics[trim={0cm 0cm 0cm 0cm},width=\columnwidth]{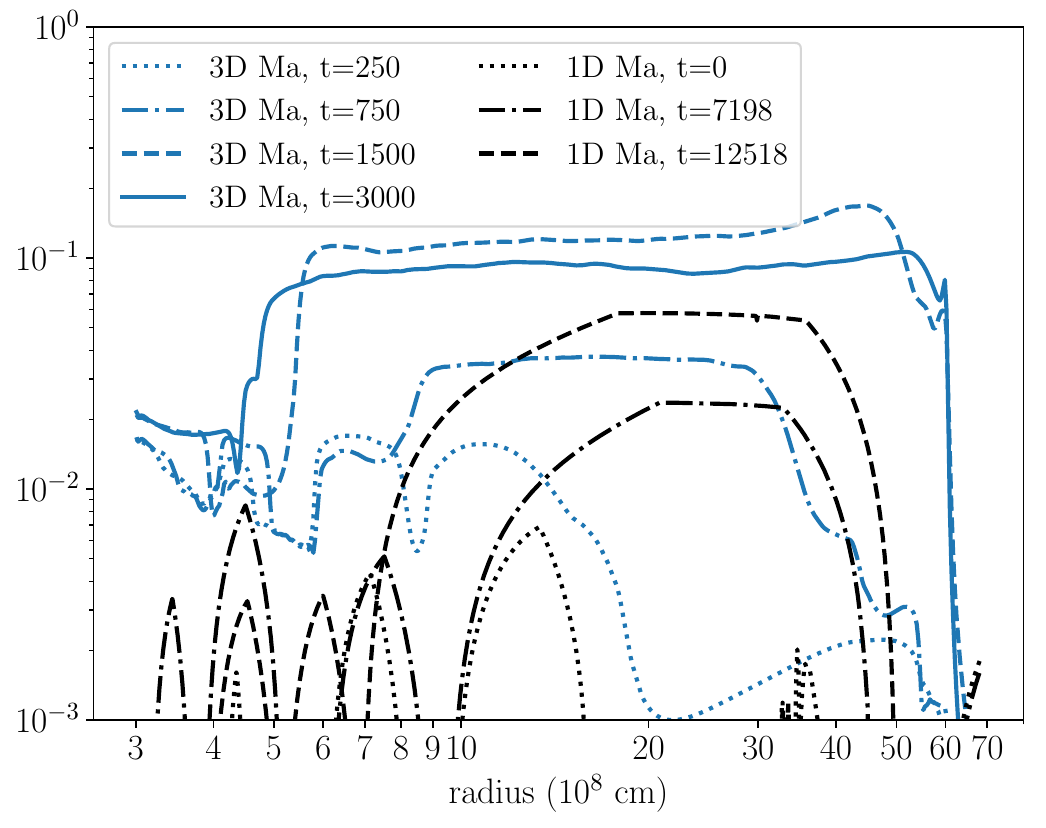}
\caption{Angularly averaged radial profiles of the Mach number Ma, comparison between 1D \texttt{MESA} model (black) and 3D model \texttt{a360n1024} (blue), at different times throughout the simulations.}\label{fig:mach}
\end{figure}

Table \ref{tab:2} shows some key quantities for each of the burning shells of \texttt{a360n1024}, at four time-steps chosen to represent the key stages of the simulation: the initial configuration at 250 s; right before and after the merging at 750 s and 1500 s, respectively; and the final state at 3000 s. It is not straightforward to analyse the evolution of convective shells with such different properties across time and between each other. The C and Ne shell mergers increase their convective velocity by almost one order of magnitude over time, reaching the maximum right after the merging, and their convective velocity can be up to 10 times larger than the one of the O shell. As a result, the convective turnover time spans a wide range of values across the different shells, requiring attention when choosing the time windows for a statistical analysis.
\begin{table}
\centering
\footnotesize
\caption{Properties of the \texttt{a360n1024} shell-merger simulation, for the three convective shells burning carbon, neon, and oxygen (refer to Fig.~\ref{fig:tke_shell}) at different key times in the simulation: convective velocity $v_\text{rms}$; shell size $\Delta r$; convective turnover time $\tau_\text{c}$. The carbon and neon shells share the same values after the merging.}\label{tab:2}
\begin{tabular}{ cccc}
\\
\hline
\hline\\[-0.7em]
\multicolumn{4}{c}{convective velocity $v_\text{rms}$ ($10^6$ cm s$^{-1}$)}\\
\\[-0.7em]
\multicolumn{1}{c|}{\diagbox{time}{shell}}&C&Ne&O\\
\\[-0.7em]
\hline\\[-0.7em]
\multicolumn{1}{c|}{250 s}&6.81&8.49&6.68\\
\multicolumn{1}{c|}{750 s}&15.2&7.33&8.41\\
\multicolumn{1}{c|}{1500 s}&\multicolumn{2}{c}{56.4}&5.26\\
\multicolumn{1}{c|}{3000 s}&\multicolumn{2}{c}{41.3}&$-$\\
\\[-0.7em]\hline
\hline\\[-0.7em]
\multicolumn{4}{c}{shell size $\Delta r$ ($10^8$ cm)}\\
\\[-0.7em]
\multicolumn{1}{c|}{\diagbox{time}{shell}}&C&Ne&O\\
\\[-0.7em]
\hline\\[-0.7em]
\multicolumn{1}{c|}{250 s}&5.12&2.28&0.56\\
\multicolumn{1}{c|}{750 s}&25.7&1.04&0.85\\
\multicolumn{1}{c|}{1500 s}&\multicolumn{2}{c}{46.0}&0.44\\
\multicolumn{1}{c|}{3000 s}&\multicolumn{2}{c}{52.0}&$-$\\
\\[-0.7em]\hline
\hline\\[-0.7em]
\multicolumn{4}{c}{convective turnover time $\tau_\text{c}$ (s)}\\
\\[-0.7em]
\multicolumn{1}{c|}{\diagbox{time}{shell}}&C&Ne&O\\
\\[-0.7em]
\hline\\[-0.7em]
\multicolumn{1}{c|}{250 s}&150&53.7&16.8\\
\multicolumn{1}{c|}{750 s}&338&28.4&20.2\\
\multicolumn{1}{c|}{1500 s}&\multicolumn{2}{c}{163}&16.7\\
\multicolumn{1}{c|}{3000 s}&\multicolumn{2}{c}{252}&$-$\\
\\[-0.7em]\hline
\end{tabular}
\end{table}
\\A detailed analysis of the velocity profiles in the simulations is presented in the comprehensive plots of Fig.~\ref{fig:velocities}, showing a comparison between the 1D and 3D simulations at three different time-steps. In particular, in 1D $v_\text{con}$ is the mixing-length-theory velocity, and $v_\text{exp}$ is the radial expansion velocity; in 3D $v_\text{con}$ is the angularly averaged root-mean-square velocity, and $v_\text{exp}$ is Favre average of the radial velocity (see next section for definitions). Overall, we can recognise the peaks in $v_\text{con}$ that indicate the same convective shells between 1D and 3D, but the shape and magnitude of the curves are intrinsically different. The \texttt{MESA} mixing-length-theory velocity (dashed, black) does not present the plateaus that characterise a well-mixed zone, and $v_\text{con}$ in the 1D shells can be up to 10 times smaller than predicted by the 3D (solid, black), whose plateaus are more regular but slightly affected by the fast nuclear burning. The larger $v_\text{con}$ is linked to the shorter time-scale of the 3D simulations compared to the original 1D model. Furthermore, the expansion velocity $v_\text{exp}$ provides insight into what is happening to the layers. Overall, $v_\text{exp}$ is always orders of magnitude lower than $v_\text{con}$, so expansion is always negligible over the time-scale of the simulation. In 1D (dashed, blue when positive, orange when negative), a small initial contraction gives way to an increasing expansion of the layers during the later phases, while in 3D (solid, blue) a small expansion is present since the initial phases but still negligible compared to the convective velocity. The key point that Fig.~\ref{fig:velocities} shows is how underestimated the convective velocity is in 1D compared to its equivalent in 3D.
\\With the aim of providing a quantitative evaluation, we show in Fig.~\ref{fig:mach} the Mach number Ma of the 1D and 3D simulations. This is a variable that largely changes across radius and over time, but we can see that Ma peaks around $10^{-2}$ at the beginning of both 1D and 3D simulations, and progressively increases throughout the merging event, reaching $5\times 10^{-2}$ in the 1D and $1\times 10^{-1}$ in the 3D shell mergers. This shows once again the wide range of regimes occurring in the simulations. As previously noted, after the shell merging in 3D the Mach number slightly decreases due to the weakening convection. 
\begin{figure*}
\centering
\footnotesize
\includegraphics[trim={0cm 0cm 0cm 0cm},width=\textwidth]{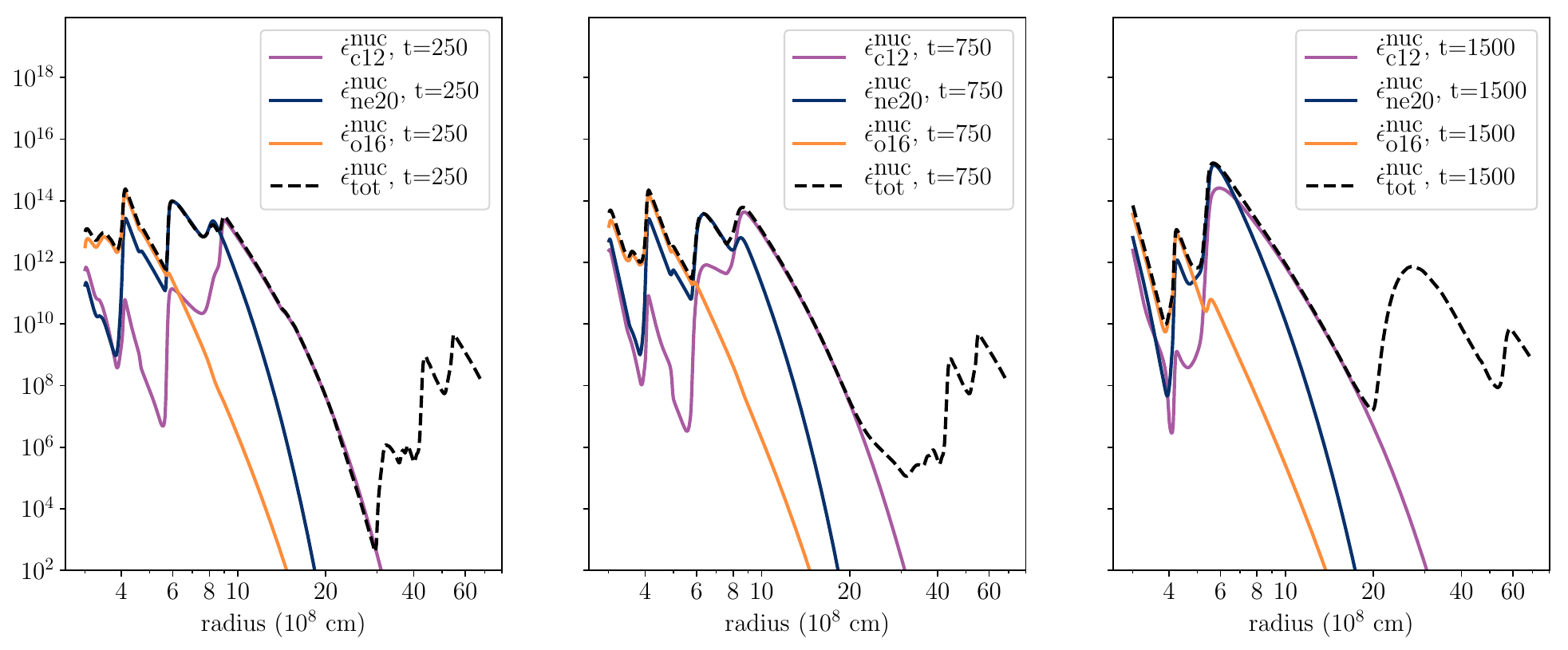}
\caption{Angularly averaged radial profiles of the nuclear energy generation rate (erg g$^{-1}$ s$^{-1}$) for C-burning ($\dot{\epsilon}^{\text{nuc}}_{\text{c12}}$), Ne-burning ($\dot{\epsilon}^{\text{nuc}}_{\text{ne20}}$), O-burning ($\dot{\epsilon}^{\text{nuc}}_{\text{o16}}$), and the total burning ($\dot{\epsilon}^{\text{nuc}}_{\text{tot}}$), from model \texttt{a360n1024}, taken at different times throughout the simulation: initial conditions (left), before (centre) and after (right) the shell merging event. }\label{fig:enuc}
\end{figure*}
\\For a better understanding of the evolution of the shells, we also present in Fig.~\ref{fig:enuc} the time evolution of different nuclear energy generation rate profiles from model \texttt{a360n1024}. The total nuclear energy is an output of the simulation, while the rates for individual reactions have been recomputed. In particular, carbon burning is reproduced by the rate of \cite{1972ApJ...176..699A}:
\begin{equation}
\dot{\epsilon}^{\text{nuc}}_{\text{c12}}=4.8\times 10^{18}\;Y^2_\text{c12}\;\rho\;\lambda_\text{c12,c12}
\end{equation}
with $Y_\text{c12}=X_\text{c12}/12$ and $\lambda_\text{c12,c12}$ the reaction rate of $\car(\car,\gamma)\magn$ from \cite{CAUGHLAN1988283}. Oxygen burning is represented by the reaction rate from \cite{1974ApJ...194..373A}:
\begin{equation}
\dot{\epsilon}^{\text{nuc}}_{\text{o16}}=8.0\times 10^{18}\;Y^2_\text{o16}\;\rho\;\lambda_\text{o16,o16}
\end{equation}
with $Y_\text{o16}=X_\text{o16}/16$ and $\lambda_\text{o16,o16}$ the reaction rate of $\oxy(\oxy,\gamma)^{32}$S from \cite{CAUGHLAN1988283}. On the other hand, neon burning is a more complex set of reactions that can be expressed with the rate of \cite{1974ApJ...193..169A}:
\begin{equation}
\begin{aligned}
&\dot{\epsilon}^{\text{nuc}}_{\text{ne20}}=4.4\times 10^{18}\;Y_\text{ne20}\;\lambda_\text{ne20}^{\gamma\alpha}\left[-7.20\left(1-A\right)+8.20B+8.35C\right]\\
&A=\dfrac{Y_\text{o16}\;\lambda_\text{o16}^{\alpha\gamma}}{\xi};\quad B=\dfrac{Y_\text{ne20}\;\lambda_\text{ne20}^{\alpha\gamma}}{\xi};\quad C=\dfrac{Y_\text{mg24}\;\lambda_\text{mg24}^{\alpha\gamma}}{\xi}\\
&\xi=Y_\text{o16}\;\lambda_\text{o16}^{\alpha\gamma}+Y_\text{ne20}\;\lambda_\text{ne20}^{\alpha\gamma}+Y_\text{mg24}\;\lambda_\text{mg24}^{\alpha\gamma}
\end{aligned}
\end{equation}
with $Y_\text{ne20}=X_\text{ne20}/20$, $Y_\text{mg24}=X_\text{mg24}/24$, and $ \lambda_\text{o16}^{\alpha\gamma}$, $\lambda_\text{ne20}^{\gamma\alpha}$, $\lambda_\text{ne20}^{\alpha\gamma}$, $\lambda_\text{mg24}^{\alpha\gamma}$ the reaction rates of $\oxy\left(\alpha,\gamma\right)$$\neo$, $\neo\left(\gamma,\alpha\right)$$\oxy$, $\neo\left(\alpha,\gamma\right)$$\magn$, $\magn\left(\alpha,\gamma\right)$$\sil$ respectively, taken from \cite{CAUGHLAN1988283}. \\
From Fig.~\ref{fig:enuc}, we can see that the three convective shells present before the merging correspond to an equal number of peaks in energy generation (at 4, 6 and $9\times10^8$ cm). The shells are fuelled by specific reactions that dominate each environment: the shell around $4\times10^8$ cm is always dominated by O-burning throughout the simulation; the two central burning shells up to the merging are fueled by Ne- and C-burning, as expected. After 1200 s, the two outermost peaks, i.e.\ the C and Ne-burning shells, merge into a single larger peak around $6\times10^8$ cm. The merged shell presents a combination of Ne- and C-burning, specifically the bottom of the shell is fueled by Ne-burning, but above $6.8\times10^8$ cm C-burning becomes dominant. At the same time, the O-burning shell reduces its energy release until it completely disappears, as we already discussed above. Additionally, some nuclear burning starts taking place also in the outer layers of the shell merger ($r>20\times10^8$ cm); this is due to the entrainment of some helium from the upper layers at $r>40\times 10^8$ cm (see Fig.~\ref{fig:shell_isorad}) and the consequent He-burning, especially the $\alpha$-capture on the abundant oxygen present, according to $\oxy(\alpha,\gamma)\neo$. The general picture is similar to what described in the simulation of \cite{Mocak2018}, where different burning regions are present within the same convective shell. Here in the shell merger we identify three distinct burning regions within the same shell: Ne-, C- and He-burning.
\\Finally, to further study the velocity fields and also highlight the differences in geometry between the two 3D simulations, we computed the power spectra of the kinetic energy with 2D Fourier transforms. We followed here the same procedure as described in \cite{2019Cristini}, \cite{Andrassy2022} and applied to spherical geometry in \cite{2023Rizzuti}. Summarizing, we fix the radius $r=1\times10^9$ cm inside the shell merging event, and compute the 2D Fourier transform of the velocity magnitude as a function of the angular coordinates $\theta$ and $\phi$:
\begin{equation}
\hat{v}_\text{rms}(k_\theta,k_\phi)=\dfrac{1}{N_\theta N_\phi}\sum_{n_\theta=0}^{N_\theta-1}\sum_{n_\phi=0}^{N_\phi-1}v_\text{rms}(\theta,\phi)e^{-i2\pi\left(\dfrac{k_\theta n_\theta}{N_\theta}+\dfrac{k_\phi n_\phi}{N_\phi}\right)}
\end{equation}
with $N_\theta,N_\phi$ the numerical resolution, $n_\theta,n_\phi$ the cell numbers, and $k_\theta,k_\phi$ the wave numbers spanning the range:
\begin{equation}
   \begin{aligned}
    &    k_\theta = \begin{cases}
        i, & \text{if\quad} 0 \le i < N_\theta/2 \\
        i-N_\theta, & \text{if\quad} N_\theta/2 \le i < N_\theta
    \end{cases}\\
    &    k_\phi = \begin{cases}
        j, & \text{if\quad} 0 \le j < N_\phi/2 \\
        j-N_\phi, & \text{if\quad} N_\phi/2 \le j < N_\phi
    \end{cases}
\end{aligned} 
\end{equation}
From this we define the specific kinetic energy $\dfrac{1}{2}|\hat{v}_\text{rms}|^{2}$, function of the wave number $k=\sqrt{k_\theta^2+k_\phi^2}$. Since the radius has been chosen to fall within the shell merger, all spectra have been averaged over one convective turnover of the C-burning shell at that moment, using the values listed in Table~\ref{tab:2}. The resulting spectra are shown in Fig.~\ref{fig:spec_merge}, for simulations \texttt{a360n1024} (top panel) and \texttt{a60n256} (bottom panel). The spectra with different geometry have a different extent in space, since \texttt{a360n1024} can reach higher $k$ compared to \texttt{a60n256} due to its larger number of cells, and larger scales $x$ due to its larger domain. \\
The ‘Kolmogorov theory’ \citep{1941DoSSR..30..301K} states that for homogeneous isotropic turbulence the rate of energy dissipation is independent of the scale and of the type of dissipative process, therefore the kinetic energy scales as $E_\text{K}\sim v^2_\text{rms}\sim k^{-5/3}$. In stellar simulations, convection is isotropic throughout the so-called ‘inertial range’, deviating from it at the smallest scales due to dissipative effects, and at the largest ones because fluid stops being isotropic. All spectra in Fig.~\ref{fig:spec_merge} follow the expected Kolmogorov scaling at most scales, but \texttt{a60n256} does so for a larger real range ($4\times 10^7$ - $4\times 10^8$ cm) compared to \texttt{a360n1024} ($6\times 10^7$ - $4\times 10^8$ cm) due to its higher local resolution, which induces dissipation at smaller spatial scales. \\
It can also be noted that some strong absorption frequencies are present in the spectra of \texttt{a360n1024}, always with approximately the same magnitude and at the same location, but they are not present in \texttt{a60n256}. These three frequencies correspond to the combinations where ($k_\theta,k_\phi$) assume the values of 2 or 6. The fact that they are constant in time, and only appear in \texttt{a360n1024} but not in \texttt{a60n256}, indicates that they are an effect of the geometry of the simulation, which in the case of \texttt{a360n1024} is close to but not exactly a full sphere. Without performing a spherical harmonic decomposition, which would be challenging and artificial in this context, it is difficult to derive any additional information about the impact of the geometry on the velocity field.
\begin{figure}
\centering
\footnotesize
\includegraphics[trim={0cm 0cm 0cm 0cm},width=\columnwidth]{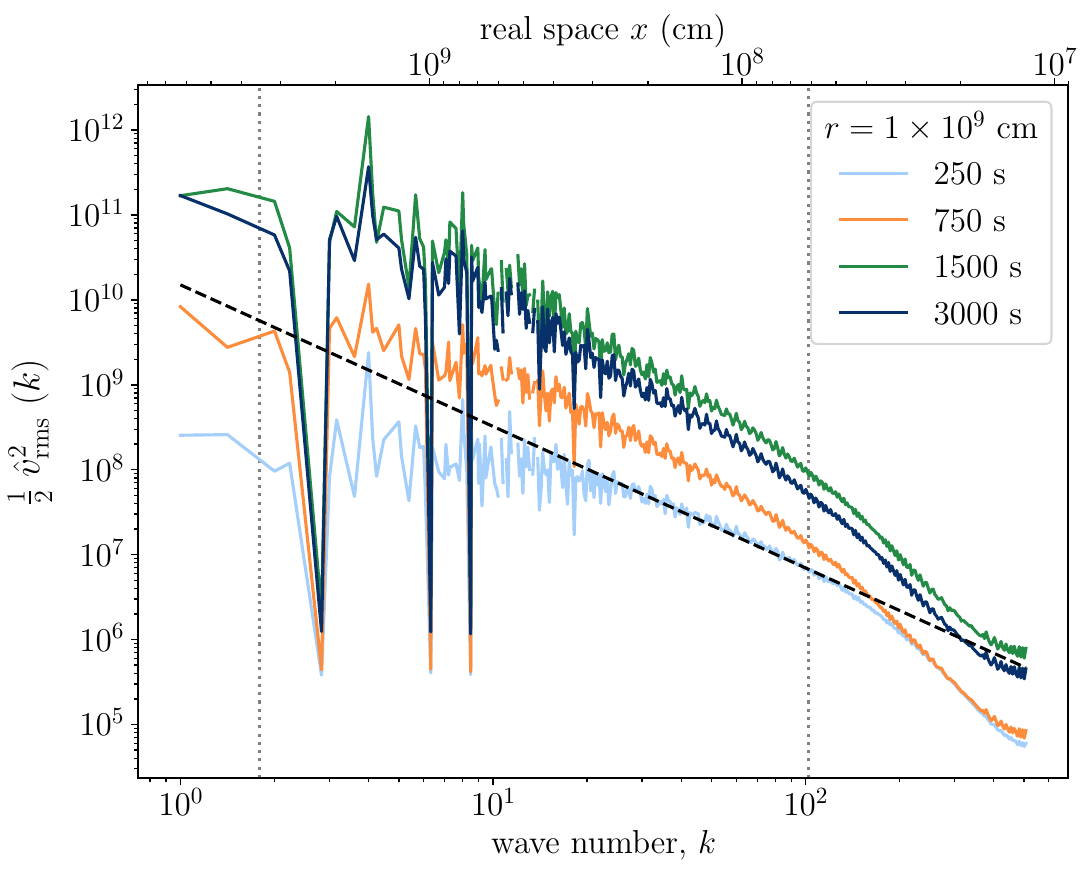}\put(-117,160){\large\texttt{a360n1024}}\\
\includegraphics[trim={0cm 0cm 0cm 0cm},width=\columnwidth]{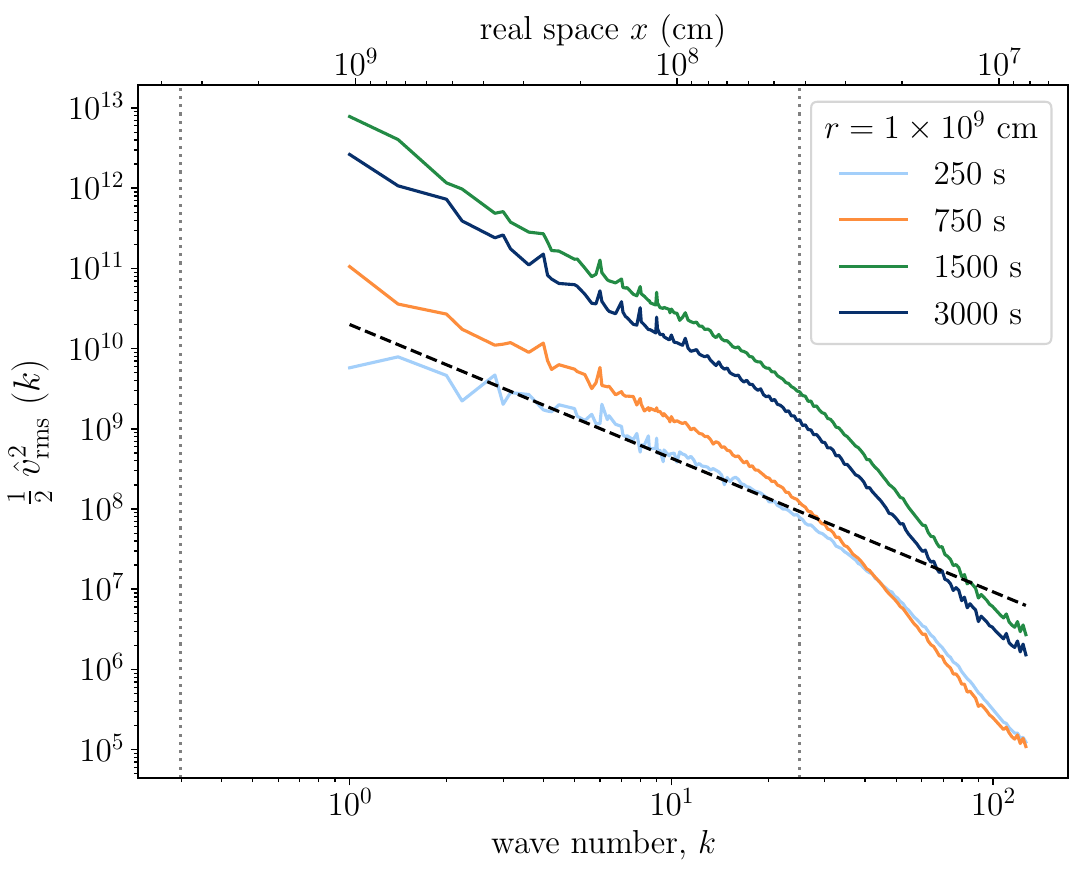}\put(-108,160){\large\texttt{a60n256}}
\caption{Specific kinetic energy spectra as function of the wave number $k$ and the real space $x$, taken inside the shell merger, for \texttt{a360n1024} (top panel) and \texttt{a60n256} (bottom panel), at different times throughout the simulations. The dashed black line is the Kolmogorov scaling $k^{-5/3}$; the vertical dotted line on the left is the average radial size of the shell merger, the one on the right is the size of 10 cells.}\label{fig:spec_merge} 
\end{figure}
\\Even though the Kolmogorov scaling is commonly invoked to interpret convection in stellar hydrodynamic simulations, we must recall that it is not conclusively established whether this choice is accurate or not in this context \citep[see e.g.][]{PhysRevLett.69.605}. For example, for stably stratified turbulence an alternative theory has been derived by \cite{Bolgiano} and \cite{Obukhov}, finding that kinetic energy scales as $E_\text{K}\sim k^{-11/5}$ (Bolgiano-Obukhov scaling) in such environment. Although it is expected that a crossover scale exists (the Bolgiano length) marking the transition between the two scalings, this has not yet been unequivocally identified either experimentally or numerically (see \citealp{Lohse} for more details).

\subsection{Evolution of the chemical composition}\label{sec:3.3}

\begin{figure*}
\centering
\footnotesize
\includegraphics[trim={0cm 0cm 0cm 0cm},width=\textwidth]{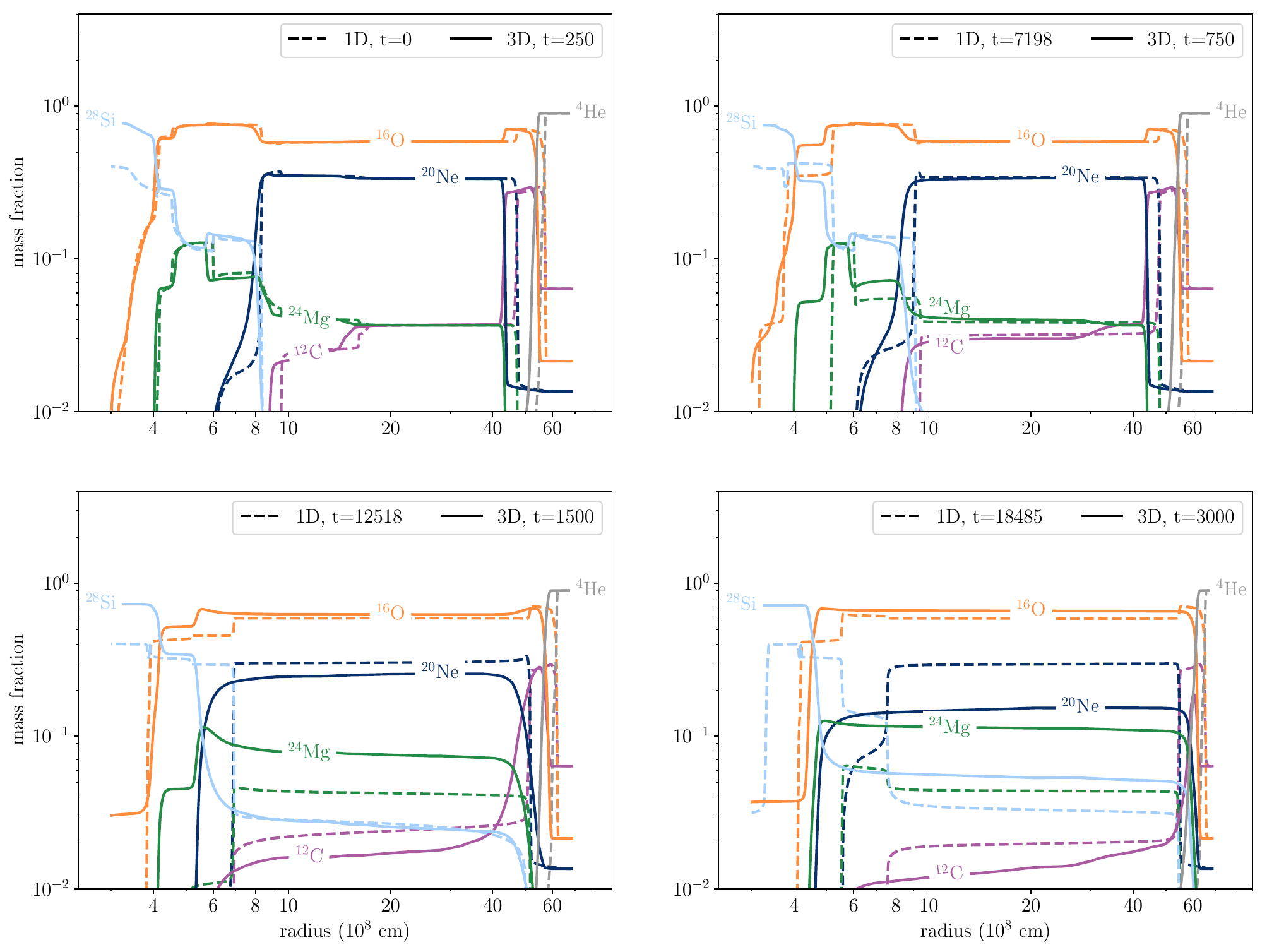}
\caption{Angularly averaged radial profiles of mass fraction for $^4$He, $\car$, $\oxy$, $\neo$, $\magn$, and $\sil$, for the 1D \texttt{MESA} model (dashed) and 3D model \texttt{a360n1024} (solid), at different times throughout the simulations: initial conditions (top, left), before (top, right) and after (bottom, left) the shell merging event, and towards the end (bottom, right).}\label{fig:shell_isorad}
\end{figure*}

\begin{figure*}
\centering
\footnotesize
\includegraphics[trim={0.75cm 4.5cm 0.75cm 4.5cm},clip,width=\textwidth]{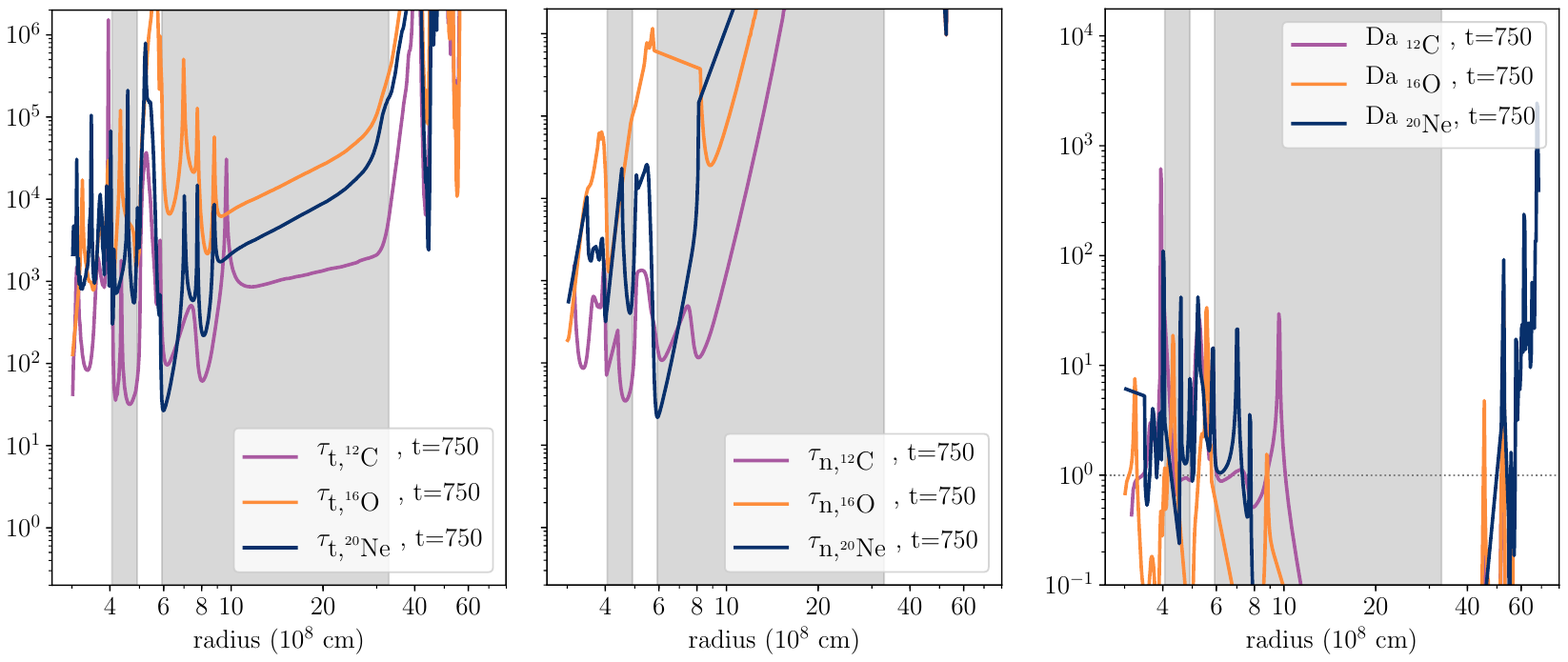}
\caption{Angularly averaged radial profiles of the transport time-scale $\tau_\text{t}$ (left), the nuclear burning time-scale $\tau_\text{n}$ (centre), and the ratio between the two (Damköhler number Da, right), for the 3D model \texttt{a360n1024} at 750 s, averaged over 333 s; the convective shells are shaded.} \label{fig:timescales}
\end{figure*}

We focus now on the analysis of the chemical abundances in the fluid of simulation \texttt{a360n1024} and their evolution over time. In this environment of shell mergers and multiple convective regions, it can be very challenging to disentangle the different contributions coming from convective mixing, nuclear burning and entrainment. In this section, we make use of a mean-field statistical analysis tool, the Reynolds-averaged Navier-Stokes (RANS) open-source code \texttt{RANSX}\footnote{\url{https://github.com/mmicromegas/ransX}}, developed for hydrodynamic simulations in spherical geometry by \cite{2014arXiv1401.5176M, Mocak2018}. Summarizing, in the RANS framework the Reynolds average (time and angular averaging) of a quantity $q$ on a spherical shell at radius $r$ is
\begin{equation}
\overline{q}(r)=\dfrac{1}{T\Delta\Omega}\int_0^T \int_{\Delta\Omega} q(r,\theta,\phi,t)\ \mathrm{d} t\ \mathrm{d}\Omega
\end{equation}
with $\mathrm{d}\Omega=\sin\theta\mathrm{d}\theta\mathrm{d}\phi$ the solid angle element, $T$ the time window, $\Delta\Omega$ the solid angle of the shell. Then, the Favre average (density-weighted average) is defined as $\widetilde{q}=\overline{\rho q}/\overline{\rho}$. Therefore, the field decomposition is done as $q=\overline{q}+q'$ or $q=\widetilde{q}+q''$, with $q',q''$ being the fluctuations of quantity $q$.\\
We present in Fig.~\ref{fig:shell_isorad} the radial distribution of some key isotopes at different time-steps, focusing on the difference between the 1D and the 3D evolution. Starting from very similar initial conditions (top, left panel), we can easily identify before the merging (top, right) the three initial convective shells: the O-burning one indicated by the plateau between 4 - $5\times10^8$ cm; the C-burning one above $9\times 10^8$ cm; and in between the Ne-burning one shown by the gradient in $\neo$ between 6 - $8\times10^8$ cm, because neon is consumed faster than it is mixed. After the merging (bottom, left) and towards the end of the simulations (bottom, right), a single large plateau is dominating most of the domain, progressively extending inwards due to entrainment, and diminishing the abundance of $\car$ and $\neo$ due to their burning, while producing $\oxy$, $\magn$ and $\sil$ as a result. Entrainment of $^4$He and $\car$ from the rich upper layers ($r>40\times 10^8$ cm) is also noticeable, contributing to the secondary peak in energy generation seen in Fig.~\ref{fig:enuc}. The comparison between 1D and 3D also yields interesting results: by the end of the simulations, not only the 3D merged shell has a larger size, but most importantly the burning has been much more efficient, completely changing the composition and consuming more $\car$ and $\neo$ while producing more $\oxy$, $\magn$ and $\sil$ compared to the 1D. Additionally, the burning time-scale is so much shorter than the mixing one that very often instead of plateaus the abundances present gradients in composition.
\\This point can be made clearer by calculating the characteristic time-scales in simulation \texttt{a360n1024}. Specifically, making use of the RANS framework, we define here \citep[see][]{Mocak2018} the nuclear burning time-scale for an isotope $i$:
\begin{equation}
\tau_{\text{n,}i}=\widetilde{X_i}/ \widetilde{\dot{X}_i}
\end{equation}
where $\dot{X}_i$ is the rate of change of isotope $i$; and the transport time-scale for isotope $i$:
\begin{equation}
\tau_{\text{t,}i}=\widetilde{X_i}/ \left( \nabla_x f_i / \overline{\rho} \right)
\end{equation}
where $f_i=\overline{\rho}\ \widetilde{X''_i v''_\text{r}}$ is the turbulent flux of isotope $i$. Finally, we also define the Damköhler number Da$_i=\tau_{\text{t,}i}/\tau_{\text{n,}i}$ as the ratio of the two time-scales. The meaning of Da is that when Da $<1$ convective mixing dominates over nuclear burning, so the convective region is always well mixed as it is the case in hydrostatic stellar burning; when instead Da $\gtrsim 1$ the nuclear burning is proceeding faster than mixing, so we see gradients in the chemical composition such as in Fig.~\ref{fig:shell_isorad}. The time-scales defined above can be computed for each isotope $i$ in the simulation, but of course the dominant time-scales (i.e.\ the shortest ones) are those of the elements that are burnt or mixed the fastest, which vary throughout the simulation.
\\We show in Fig.~\ref{fig:timescales} the time-scales and Da number for $\car,\oxy$ and $\neo$ in \texttt{a360n1024} right before the merging, as representative of the burning in the multiple convective shells; we include a more comprehensive study in Appendix \ref{append:B}, Fig.~\ref{fig:time-timescales} at three different time-steps. The first thing that we see is that at the bottom of the convective regions (shaded in grey) there is a dip in the time-scales, corresponding to the element that is burnt or mixed the fastest in the Ne- and C-burning shells; for the O-shell instead, the element that is processed the fastest is $\car$, but this is due to its very low abundance in the shell. At all times, in the outer regions ($r>10\times10^8$ cm) the transport always strongly dominates over the nuclear burning, as we can see from the very low Da, but for the inner regions this is often not the case. The Damköhler number in the inner regions strongly fluctuates and often exceeds 1, presenting some important peaks above Da $\sim10$. Looking carefully, we can see that some of the largest peaks correspond to the bottom of the convective shells, where the burning is taking place. Some of the regions where $\text{Da}>1$ inside the convective shells are large, and they correspond to the layers where a gradient in chemical composition is maintained (compare to Fig.~\ref{fig:shell_isorad}), such as the Ne-burning region at $r=6\text{ - }7\times10^8$ cm, and the C-burning one at $r=9\text{ - }10\times10^8$ cm, before the merging. Overall, we can conclude that in this highly dynamical environment the nuclear burning is so efficient that the convective shells are rarely well mixed, especially near the burning locations. 
\begin{figure*}
\centering
\footnotesize
\includegraphics[trim={0cm 0cm 0cm 0cm},width=\textwidth]{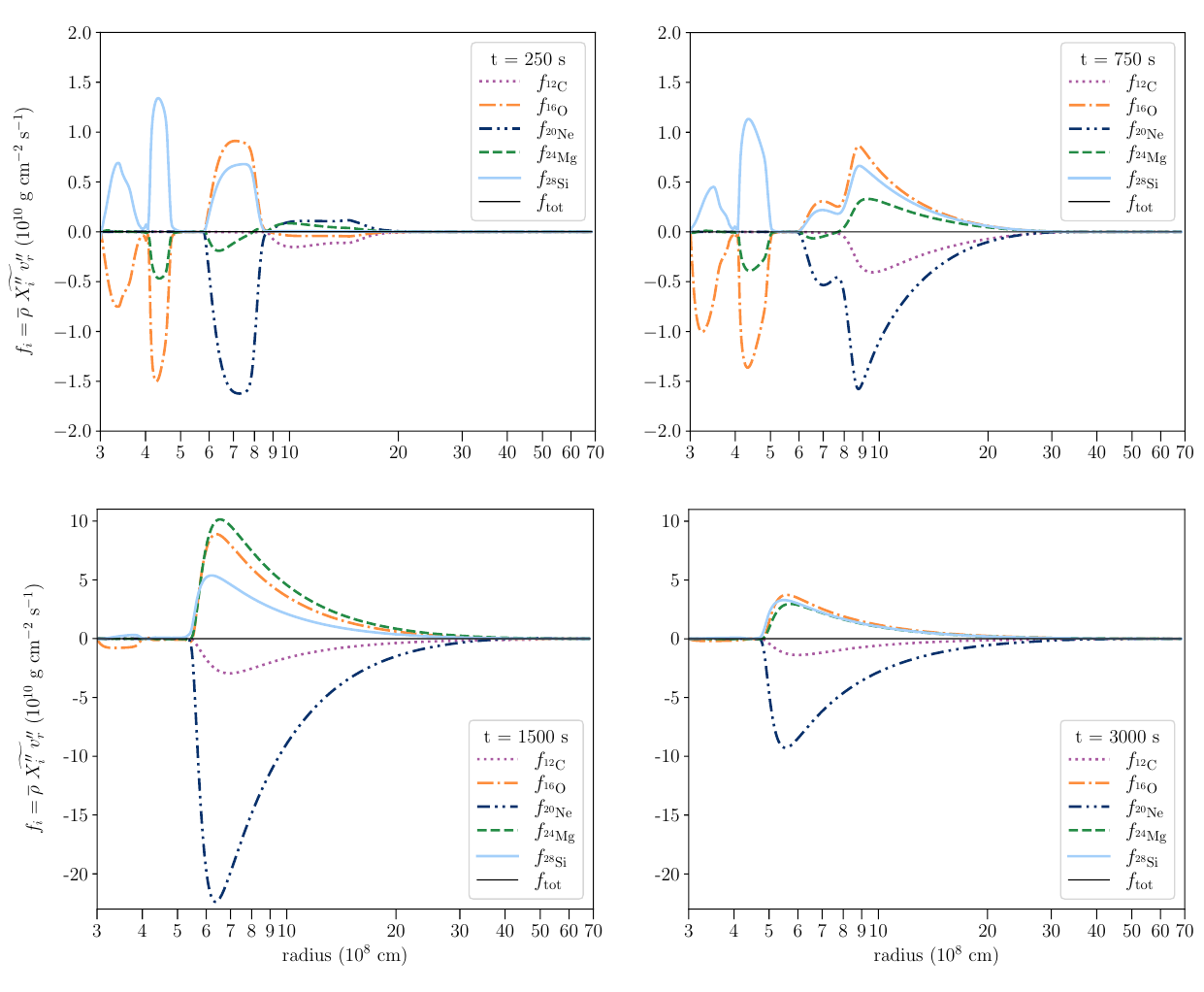}
\caption{Radial flux profiles of $\car,\ \oxy,\ \neo,\ \magn,$ and $\sil$, from simulation \texttt{a360n1024} at 250 s (top, left), 750 s (top, right), 1500 s (bottom, left) and 3000 s (bottom, right), averaged over 150 s, 333 s, 200 s and 500 s, respectively. The black line is the sum of the flux profiles for all the 12 isotopes in the network.}\label{fig:shellflux1}
\end{figure*}
\\Focusing now only on the transport mechanism, it is possible to study the radial transport of the isotopes with the mean-field radial flux that we defined above as $f_\textit{i}=\overline{\rho}\ \widetilde{X''_\textit{i} v''_\textit{r}}$ for a species $i$. In Fig.~\ref{fig:shellflux1} we show the flux profiles for the most important isotopes of simulation \texttt{a360n1024}, at the usual four time-steps. The evolution of the flux for each element is a direct result of the mixing of material that brings the fuel towards the burning regions and the ashes away from them; in this sense, the flux profiles are also a proxy of the nuclear burning occurring in each layer. Here, positive and negative values of the flux represent upward and downward transport of species, respectively. At the beginning of the simulation (top, left of Fig.~\ref{fig:shellflux1}), the four convective regions are clearly identified by the peaks and valleys in the flux, as a result of the upward and downward transport of ashes and fuel, respectively. Specifically, the two innermost shells are burning mainly $\oxy$ to produce $\sil$; the central shell between $\text{6 - }8\times 10^8$ cm is burning mostly $\neo$ to produce $\oxy$ and $\sil$; and the outermost shell is burning mostly $\car$ to produce $\neo$ (compare to Fig.~\ref{fig:tke_shell}). When the C and the Ne-burning shells merge ($t=750 \text{ - } 1500 \text{ s}$, Fig.~\ref{fig:shellflux1}), a single convective region forms, burning both $\neo$ and $\car$ to generate $\oxy$, $\magn$ and $\sil$. We also note that the magnitude of the flux increases right after the merging by an order of magnitude, as a result of the increase in radial velocity, and later with time it starts reducing again, as it is the case also for the kinetic energy (see Fig.~\ref{fig:tke_shell_time}).
\\Finally, we calculate the normalized standard deviation of the mass fraction, defined as $\sigma_{i}/\bar{X}_{i}=(\widetilde{X''_{i} X''_{i}})^{1/2}/\bar{X}_{i}$ for a species $i$. Figure \ref{fig:devshell} shows the $\sigma_{i}/\bar{X}_{i}$ profiles for the key isotopes of simulation \texttt{a360n1024} before and after the shell merging. Although it is challenging to distinguish the precise layers in these plots, they still provide an estimate of the magnitude of the chemical dispersion, which also represents the deviation from the spherical symmetry assumed in 1D models. Before the merging, the normalized dispersion can reach up to 200 per cent at the convective boundaries, but inside the convective regions it is closer to 10 - 20 per cent for $\car$ and up to a few per cent for the other isotopes. However, after the merging the peaks in normalized deviation go beyond 300 per cent, while the values inside the merged region have also increased to 30 per cent for $\car$ and $\sil$, and at least 10 per cent for $\neo$ and $\magn$. Overall, these values are very large, especially compared to the dispersion in more conventional burning phases, such as the Ne-burning shell in \cite{2023Rizzuti} (see their fig.~15). \\
To study these variations in more detail, we also show in Fig.~\ref{fig:shellcar} the fluctuations in $\car,\oxy$ and $\neo$ abundances around the mean value, in angular cross-sections taken after the shell merging. The plot shows that the fluctuations are not mixed homogeneously across the surface, but they are arranged in large-scale structures, dividing the surface in almost two distinct regions, of which one is overabundant in $\oxy$ and underabundant in $\car$ and $\neo$, reflecting the nuclear burning discussed above. This behaviour is an effect of the highly dynamical environment of the shell mergers, and it can have a significant impact on the stellar nucleosynthesis, especially in relation to the following supernova explosion, leading to large fluctuations in the different abundances within the ejected material. 
\begin{figure*}
\centering
\footnotesize
\includegraphics[trim={0cm 0cm 0cm 0cm},width=0.51\textwidth]{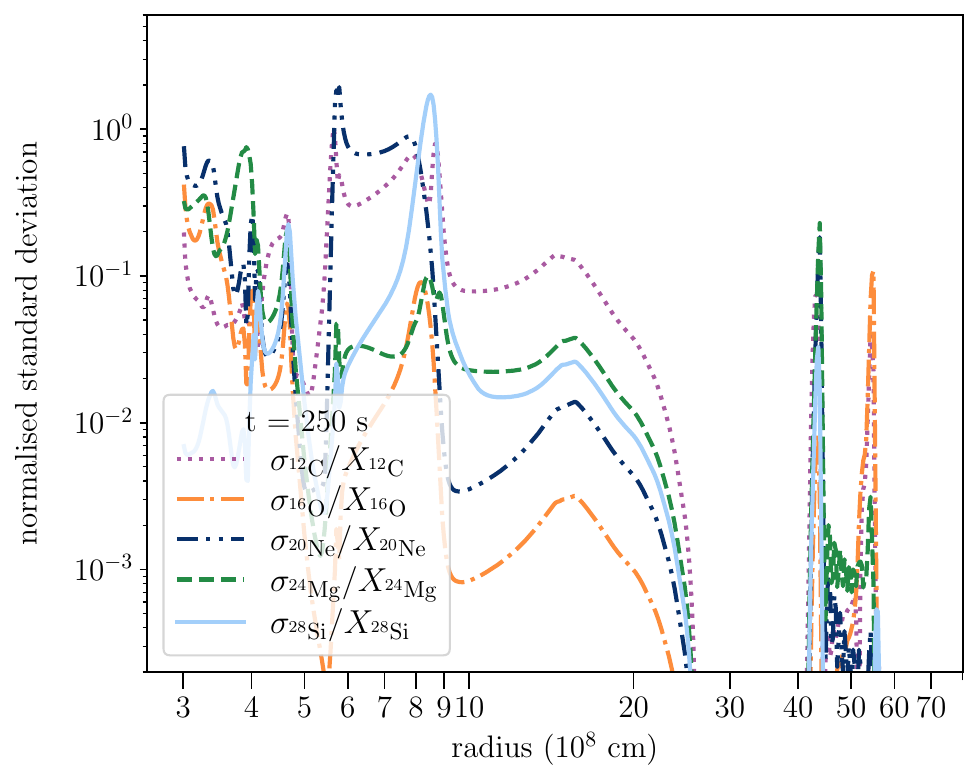}
\includegraphics[trim={1cm 0cm 0cm 0cm},clip,width=0.48\textwidth]{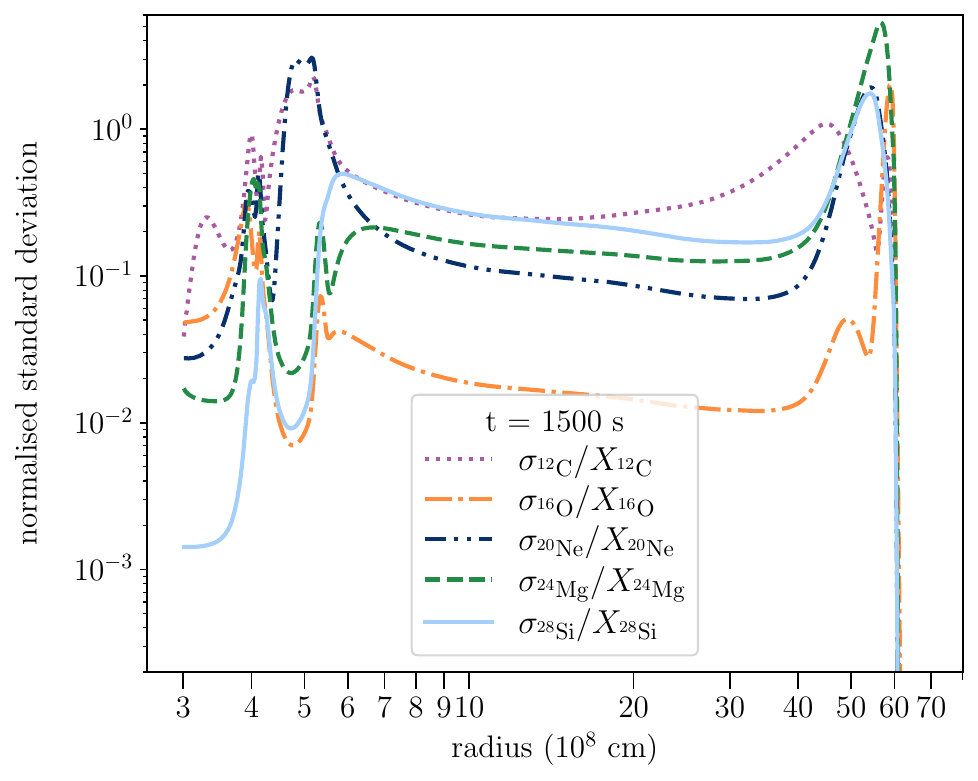}
\caption{Normalized standard deviation profiles of $\car,\ \oxy,\ \neo,\ \magn,$ and $\sil$ as a function of the stellar radius,  from simulation \texttt{a360n1024} at 250 s (left) and 1500 s (right), averaged over 150 s and 200 s, respectively.}\label{fig:devshell}
\end{figure*}
\begin{figure*}
\centering
\footnotesize
\includegraphics[trim={0cm 0cm 0cm 0cm},width=\textwidth]{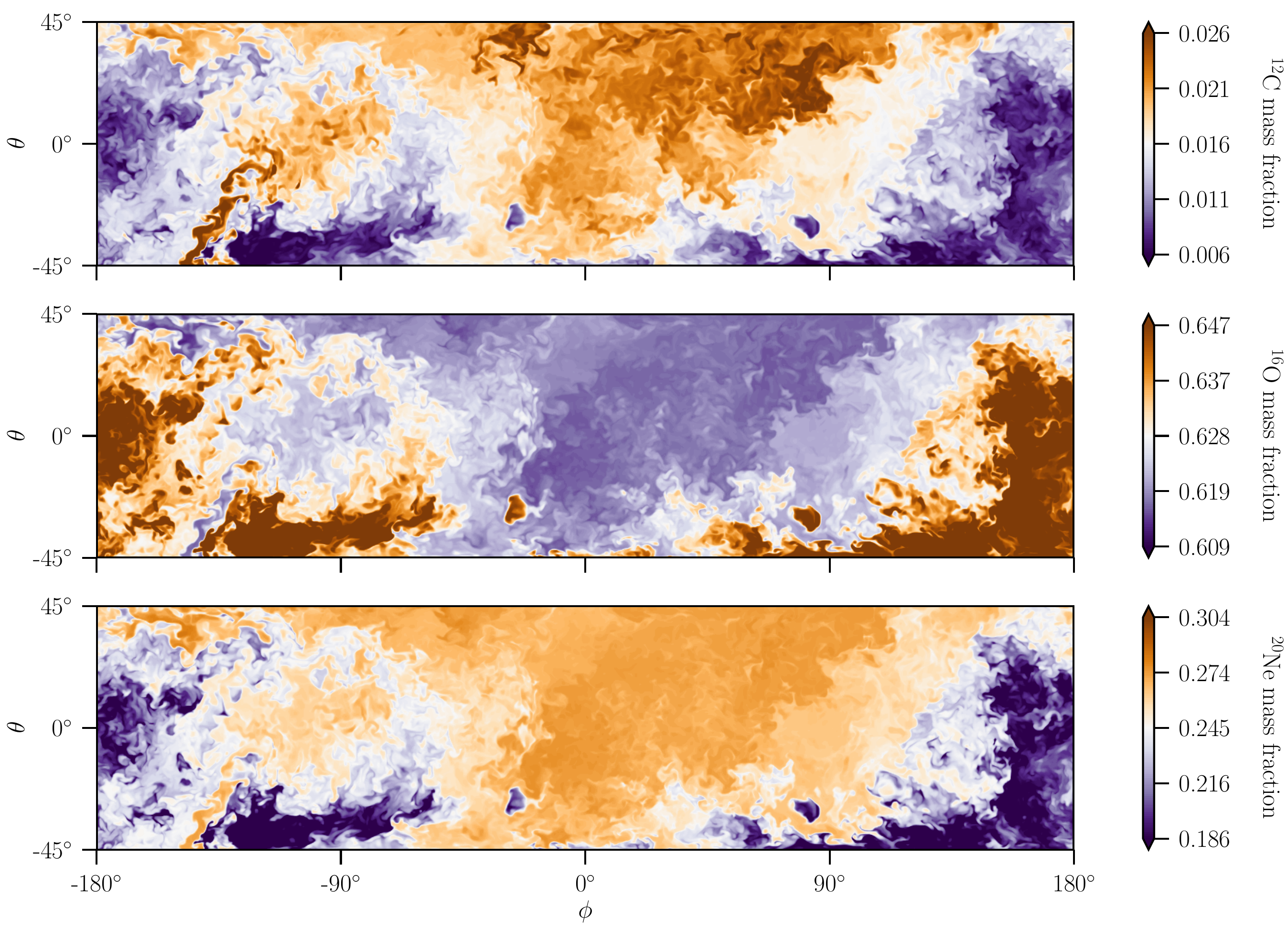}
\caption{Angular cross-sections taken at $r=1\times 10^9$ cm from simulation \texttt{a360n1024} at 1500 s, right after the shell merging, showing the fluctuations in $\car,\oxy$ and $\neo$ mass fractions in colour scale, spanning $2\sigma$ around the mean value.}\label{fig:shellcar}
\end{figure*}

\section{Conclusions}

A good understanding of the structure and evolution of massive stars is crucially important for many fields in astrophysics. This becomes particularly challenging when addressing the advanced phases of massive stars, whose internal structure is organised in concentric shells with multiple convective regions that can interact. In this paper, we have produced and analysed 3D hydrodynamic simulations of the merging of two convective shells into a single larger one, several hours before the final collapse of the star. This environment is of great interest for studying the evolution of the stellar structure towards its final phases and the nucleosynthesis paths that can be enabled. \\
The simulations have been started from initial conditions assumed from a 1D model of a 20 M$_\odot$ star that predicts the merging of multiple convective shells. We designed a nearly $4\pi$ geometry, performing a comparison with the more usual `wedge' geometry, and we ran the 3D stellar simulations at nominal luminosity (i.e.\ without altering the energy generation) including an extensive nuclear network to reproduce the multiple burning phases. The 3D simulations confirm the occurrence of a merging between the carbon-burning and the neon-burning shells as shown in the original 1D model. We analysed both the dynamics and nucleosynthesis aspects of the 3D simulations, making use of mean-field analysis tools, and focusing on the significant differences from the initial-condition 1D model. \\
Differently from the 1D model, the merging in 3D is driven by the entrainment of material from the stable regions into the convective ones, progressively eroding the stable layers that separate convective shells and making the merging possible. The time-scale for the merging is much faster in 3D than in 1D: while this can still be an effect of the new equilibrium of the stratification recomputed onto the 3D grid, we cannot ignore the fact that several recent studies \citep{2018arXiv181004659A,2023Rizzuti,2024MNRAS.531.4293G} indicate that the mixing-length-theory velocity assumed in 1D models may be actually underestimated. \\
The shell merger we simulated also proved to be an interesting site for nucleosynthesis, presenting an efficient nuclear burning of both carbon and neon within the same convective region, as also found in similar environments \citep{Mocak2018}. Due to this very energetic burning, the chemical composition and size of the convective regions after the merging in 3D are very different from their equivalent in 1D. Most importantly, the angular dispersion of species is extremely large, as a result of the highly dynamical environment, marking a net difference from the 1D model, where by definition of spherical symmetry the angular dispersion is always null. In the context of stellar nucleosynthesis, this behaviour could have a noticeable impact on the predictions of stellar yields for massive stars, affecting the structure and composition of the ejecta.\\
With this paper, we aimed to address some of the uncertainties related to the evolution of massive stars, in order to improve our understanding of the complex multi-D processes and provide insight into the simplifying assumptions included in 1D stellar models. This is particularly important considering that, given the very high computational cost required for running multi-dimensional simulations, the 1D models remain the main tools for predicting and explaining the evolution of stellar populations. It is the interplay between 1D and multi-D models that really pushes forwards our knowledge of stellar evolution. 1D models have been shown to require revision in their treatment of convection, in particular concerning the size of convective zones and the shape of convective boundaries, as discussed in recent works \citep{2018arXiv181004659A,2019Cristini,2023Rizzuti,2024A&A...683A..97A}. \\
On the other hand, hydrodynamic models have also significant potential for improvement. Recent studies have started simulating the late stages of stellar convection including self-consistent magnetic fields coming from dynamo effects (\citealp{2021MNRAS.504..636V}; \citealp{2023A&A...679A.132L}; \citealp{2023MNRAS.526.5249V}); this will put important constraints on the amount of kinetic energy and turbulent motions that the fluid can build up. Additionally, recent works are also showing that a continuously increasing number of chemical species can be now included directly into the numerical simulations (\citealp{2015ApJ...808L..21C}; \citealp{Muller_2016}; \citealp{Mocak2018}; \citealp{Yoshida_2019}), contributing to the energy release and production of new species. With the advent of larger computing facilities and new types of processing units, it will be possible to run stellar simulations with progressively increasing resolution and longer time-scales.\\
Finally, we recall that a good understanding of stellar structure and evolution has a significant impact on different fields, ranging from studies of supernova progenitors and explosion mechanisms \citep{2015MNRAS.448.2141M,2021Natur.589...29B}, through the production of accurate progenitor models as initial conditions, to predictions on the nature and physics of the remnants and the final-initial mass relation \citep{2020MNRAS.496.1967K,2021Scott}, but also the interpretation of observations and asteroseismic measurements \citep{2021A,2021NatAs...5..715P}, to nuclear physics and galactic chemical evolution. This multi-disciplinary approach will help tackle astrophysical problems that are still unsolved today, such as the red supergiant problem \citep{2009MNRAS.395.1409S} and the black hole mass gap \citep{2021ApJ...912L..31W}. 

\section*{Acknowledgements}

FR acknowledges the grant PRIN project No. 2022X4TM3H `Cosmic POT' from Ministero dell’Università e della Ricerca (MUR). RH acknowledges support from the World Premier International Research Centre Initiative (WPI Initiative), MEXT, Japan, the IReNA AccelNet Network of Networks (National Science Foundation, Grant No. OISE-1927130) and the Wolfson Foundation. WDA acknowledges support from the Theoretical Astrophysics Program (TAP) at the University of Arizona and Steward Observatory. CG has received funding from the European Research Council (ERC) under the European Union’s Horizon 2020 research and innovation program (Grant No. 833925). The University of Edinburgh is a charitable body, registered in Scotland, with Registration No. SC005336. CG, RH, and CM acknowledge ISSI, Bern, for its support in organising collaboration. This article is based upon work from the ChETEC COST Action (CA16117) and the European Union’s Horizon 2020 research and innovation programme (ChETEC-INFRA, Grant No. 101008324). The authors acknowledge the STFC DiRAC HPC Facility at Durham University, UK (Grants ST/P002293/1, ST/R002371/1, ST/R000832/1, ST/K00042X/1, ST/H008519/1, ST/ K00087X/1, and ST/K003267/1). 

\section*{Data Availability}

The data underlying this article will be shared on reasonable request to the corresponding author.



\bibliographystyle{mnras}
\bibliography{references} 




\appendix

\section{Insight into the 3D simulations}\label{append:A}
\begin{figure*}
\centering
\footnotesize
\includegraphics[trim={0cm 0cm 0cm 0cm},width=0.85\textwidth]{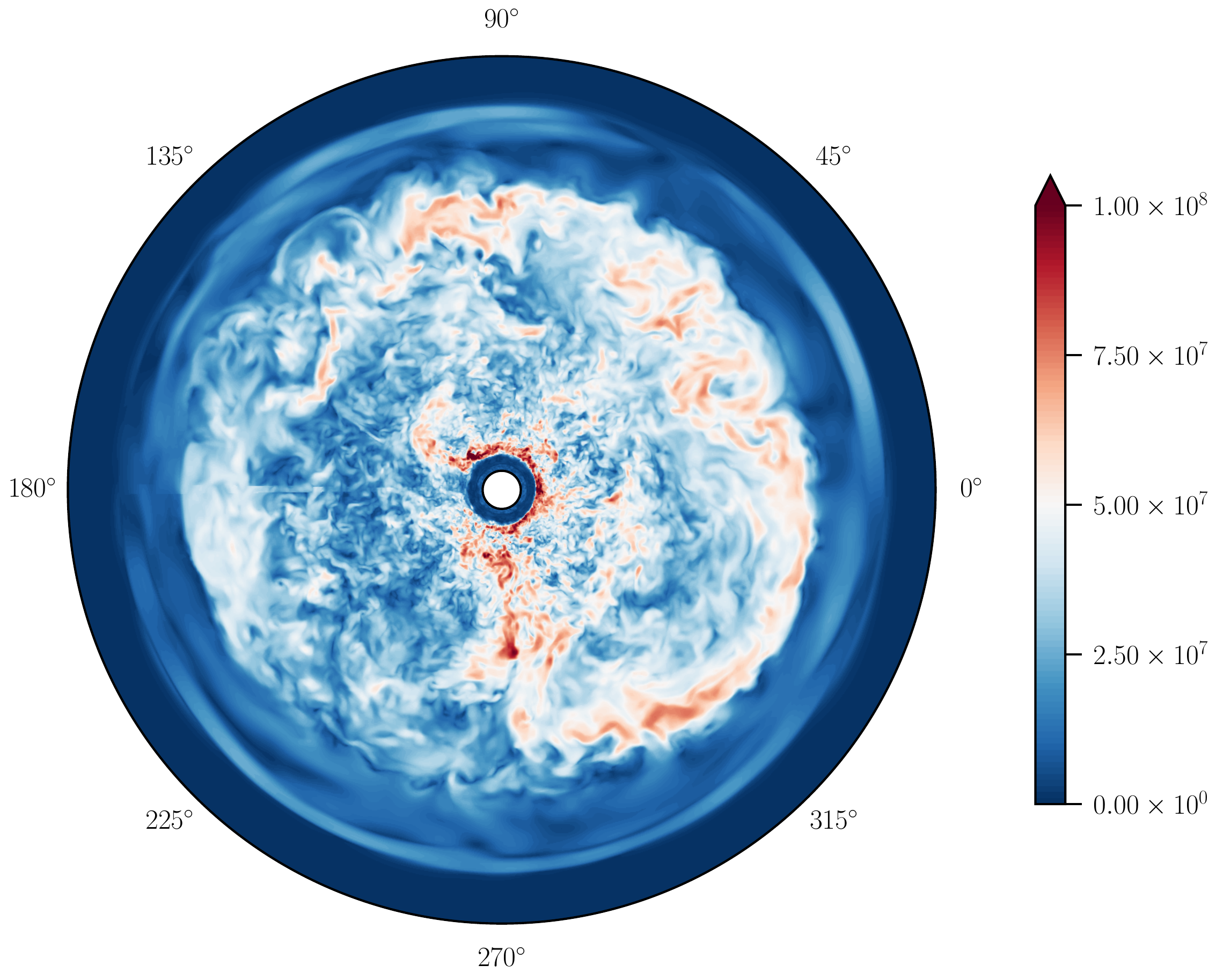}
\includegraphics[trim={2.5cm 0cm 0cm 0cm},width=0.85\textwidth]{fig/shell_90_1500s_f1024a360}
\caption{Equatorial (top panel) and longitudinal (bottom panel) cross-sections taken from \texttt{a360n1024} after 1500 seconds, with the fluid speed in colour scale in cm s$^{-1}$. The two frames show the 360$^{\circ}$ range of the $\phi$-angle (top) and two opposite sides of the 90$^{\circ}$ range of the $\theta$-angle (bottom).}\label{fig:360frame}
\end{figure*}

In order to better illustrate the structure of the 3D simulations presented in this work, and complementarily to Fig.~\ref{fig:model}, we display in Fig.~\ref{fig:360frame} two cross-sections taken from \texttt{a360n1024}, one showing the equatorial plane (top panel) and the other a longitudinal plane from two sides of the polar axis (bottom panel).

\section{Evolution of the time-scales}\label{append:B}
\begin{figure*}
\centering
\footnotesize
\includegraphics[trim={0cm 0cm 0cm 0cm},width=\textwidth]{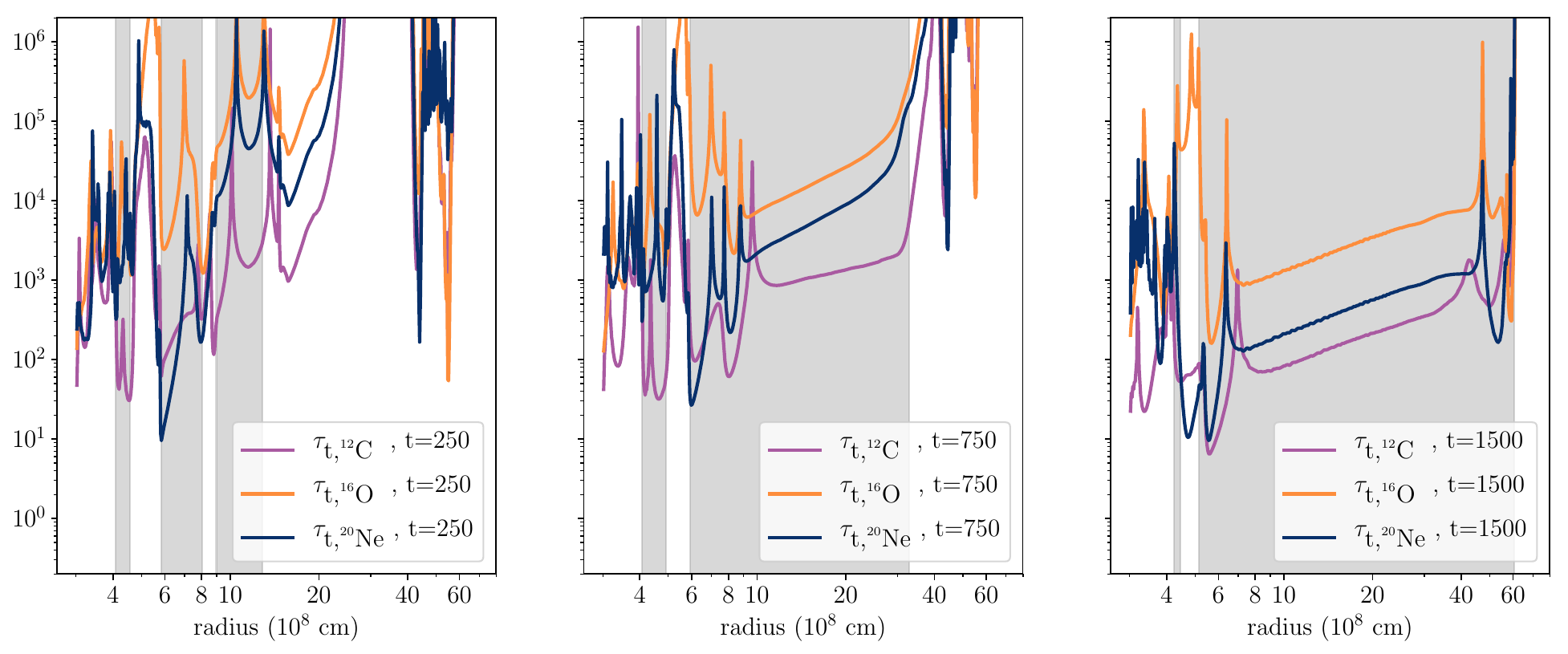}
\includegraphics[trim={0cm 0cm 0cm 0cm},width=\textwidth]{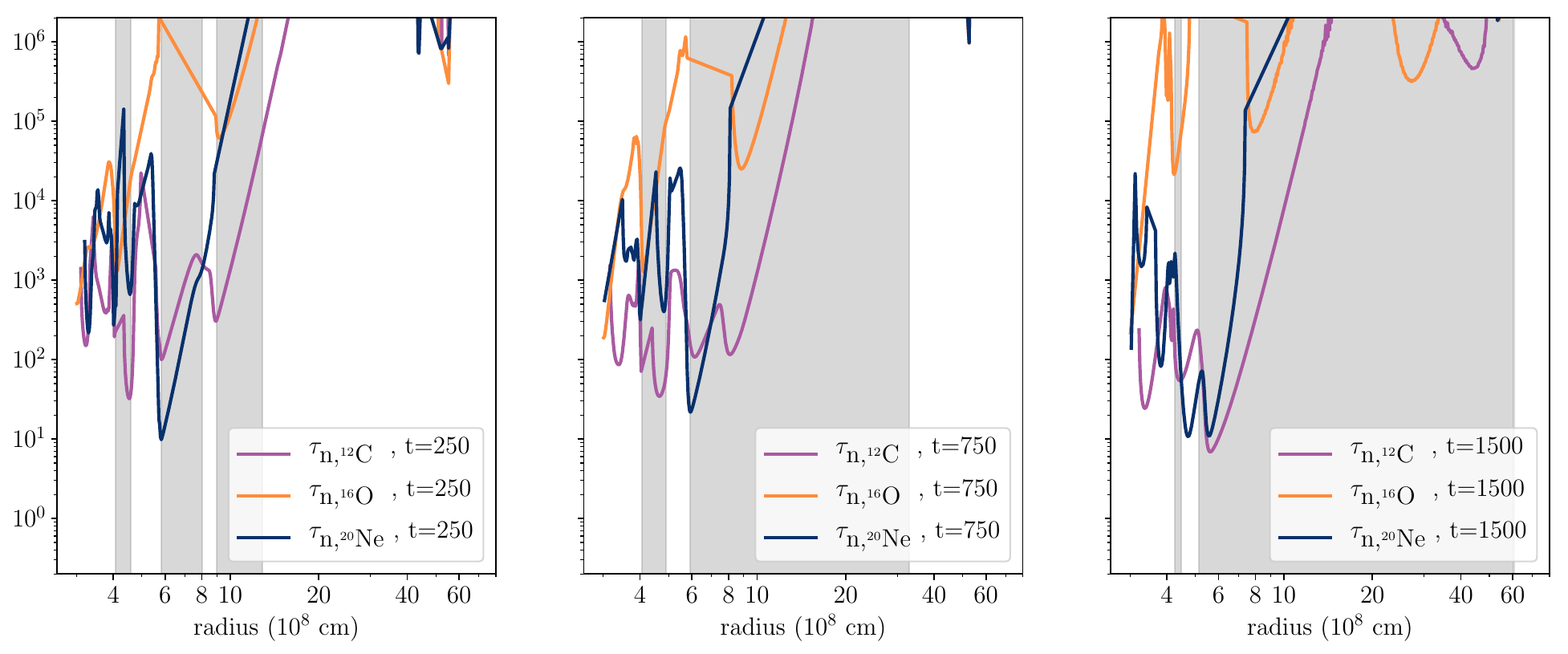}
\includegraphics[trim={0cm 0cm 0cm 0cm},width=\textwidth]{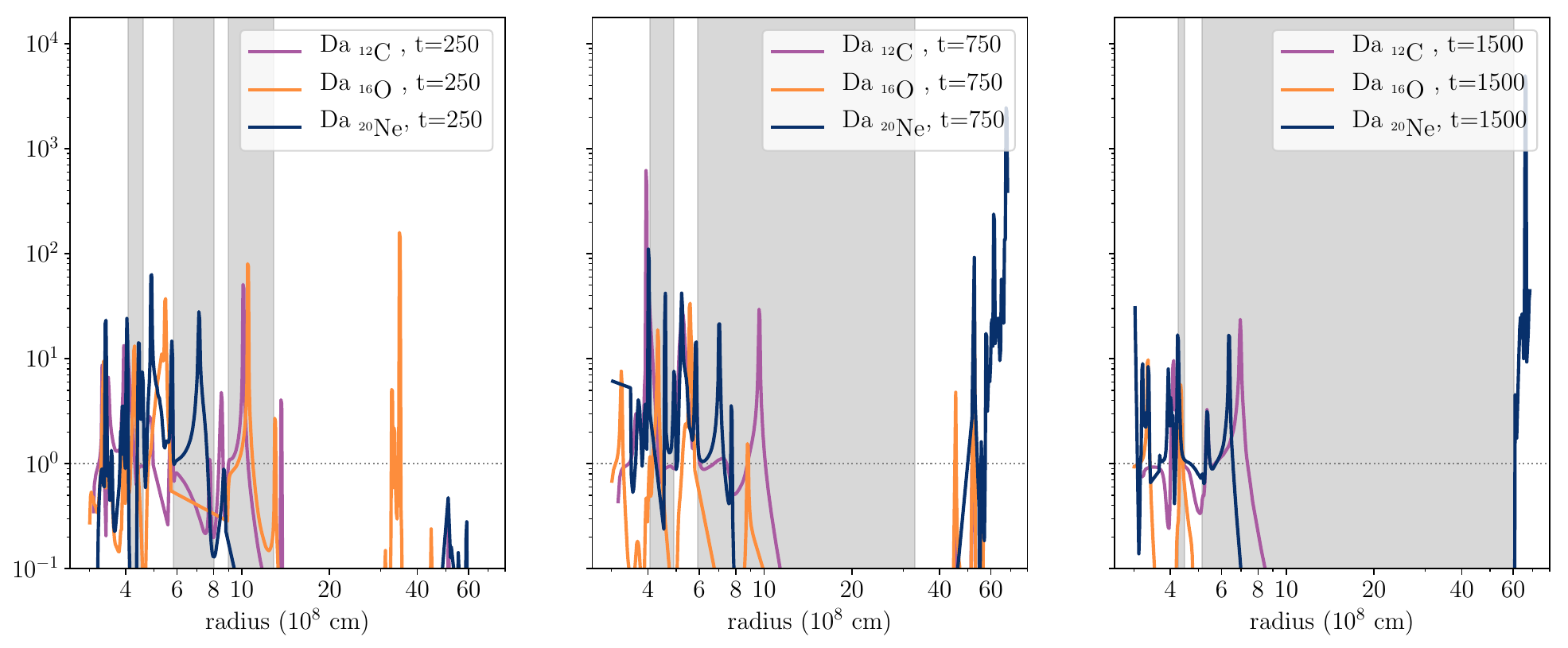}
\caption{Angularly averaged radial profiles of the transport time-scale $\tau_\text{t}$ (top panels), the nuclear burning time-scale $\tau_\text{n}$ (middle panels), and the ratio between the two (Damköhler number Da, bottom panels), for the 3D model \texttt{a360n1024}, at different times throughout the simulation: initial conditions (left), before (centre) and after (right) the shell merging event, averaged over 150 s, 333 s, and 200 s respectively; the convective shells are shaded.} \label{fig:time-timescales}
\end{figure*}

We display here the time evolution of the key time-scales in the hydrodynamic simulations, as introduced in Section \ref{sec:3.3}: the transport time-scale $\tau_\text{t}$, the nuclear burning time-scale $\tau_\text{n}$, and the ratio between the two (Damköhler number, Da), for isotopes $\car,\oxy$ and $\neo$. Complementarily to Fig.~\ref{fig:timescales}, we plot in Fig.~\ref{fig:time-timescales} the radial profiles of the time-scales and Damköhler number at three different time-steps throughout simulation \texttt{a360n1024}. 


\bsp	
\label{lastpage}
\end{document}